\documentclass[11pt]{article}

\usepackage{graphicx}
\usepackage[english]{babel}

\usepackage{amsmath}
\usepackage{slashed}

\usepackage[tight,hang,raggedright,normalsize]{subfigure}

\usepackage{picins}
\usepackage{hyperref}  %links for pdf files

\newcommand{\e}{\mathrm{e}}
\newcommand{\diff}{\mathrm{d}}

\newcommand{\Li}{\operatorname{Li}\nolimits}

\renewcommand{\Im}{\operatorname{Im}}
\renewcommand{\Re}{\operatorname{Re}}

\newcommand{\gev}{\operatorname{GeV}}

\newcommand{\lrD}{{D^{\hspace{-0.8em}
      \raisebox{0.8ex}{$\scriptstyle\leftrightarrow$}}}{}}
\newcommand{\lD}{{D^{\hspace{-0.8em}
      \raisebox{0.8ex}{$\scriptstyle\leftarrow$}}}{}}
\newcommand{\rD}{{D^{\hspace{-0.8em}
      \raisebox{0.8ex}{$\scriptstyle\rightarrow$}}}{}}

\newcommand{\slim}{\mskip 1.5mu}

\newcommand{\half}{\tfrac{1}{2}}

\newcommand{\lsim}{\raisebox{-4pt}{$
\,\stackrel{\textstyle <}{\sim}\,$}}

\begin{document}

\allowdisplaybreaks[2]

\begin{flushright}
\begin{tabular}{l}
DESY-07-032 \\
hep-ph/0703148
\end{tabular}
\end{flushright}

\vspace*{2em}

\begin{center}
\textbf{\LARGE Exclusive electroproduction of pion pairs}

\vspace*{1.4cm}

{\large
N. Warkentin$\slim{}^1$,
M. Diehl$\slim{}^2$,
D.Yu.~Ivanov$\slim{}^3$
and
A.~Sch\"afer$\slim{}^1$
}

\vspace*{0.4cm}

{\sl
$^1$ Institut f{\"u}r Theoretische Physik, Universit{\"a}t Regensburg,
     93040 Regensburg, Germany \\
$^2$ Deutsches Elektronen-Synchrotron DESY, 22603 Hamburg, Germany \\
$^3$ Sobolev Institute of Mathematics, 630090 Novosibirsk, Russia
}

\vspace*{1.8cm}

\textbf{Abstract}\\[10pt]
\parbox[t]{0.9\textwidth}{
  We investigate electroproduction of pion pairs on the nucleon in the
  framework of QCD factorization for hard exclusive processes.  We
  extend previous analyses by taking the hard-scattering coefficients
  at next-to-leading order in $\alpha_s$.  The dynamics of the
  produced pion pair is described by two-pion distribution amplitudes,
  for which we perform a detailed theoretical and phenomenological
  analysis.  In particular, we obtain constraints on these quantities
  by comparing our results with measurements of angular observables
  that are sensitive to the interference between two-pion production
  in the isoscalar and isovector channels.}

\vspace*{1cm}
\end{center}

%#####################################
%
%       INTRODUCTION
%
%#####################################

\section{Introduction}

The theoretical treatment of hard exclusive processes has been a
challenge in QCD for many years. With the advent of the generalized
parton distribution (GPD) formalism a large class of such processes,
all involving some hard scale $Q^2$, can now be treated on a firm
basis, with all non-perturbative physics described by suitable
generalized parton distributions and similar quantities.  GPDs allow
one to relate the information from many different processes within an
overall QCD description, including aspects that cannot be deduced
directly from experiment, like the transverse spatial distribution of
partons and their orbital angular momentum.  Pioneering papers of this
field are
\cite{Mueller:1998fv,Ji:1996ek,Radyushkin:1996nd,Burkardt:2000za}, and
extensive reviews are given in
\cite{Goeke:2001tz,Diehl:2003ny,Belitsky:2005qn}.

The large amount of information contained in GPDs implies that much
and diverse data is needed to reliably determine their functional form
with respect to the three variables $x$, $\xi$, and $t$ (see below).
One of the channels for which data is available is 
exclusive electroproduction of pion pairs \cite{Airapetian:2004sy}.
This process has already been studied by some of us
\cite{Lehmann-Dronke:1999aq,Lehmann-Dronke:2000xq} at leading order
(LO) in $\alpha_s$, and will be analyzed here at next-to-leading order
(NLO).  For this we can use various results
\cite{Belitsky:2001nq,Ivanov:2004zv} obtained earlier for similar
processes, which greatly simplifies our task.  Apart from GPDs, a
second non-perturbative input in the description of pion pair
production are two-pion distribution amplitudes ($2\pi$DAs),
introduced in \cite{Mueller:1998fv,Diehl:1998dk}.  Building on earlier
work in \cite{Polyakov:1998ze,Kivel:1999sd,Diehl:2000uv} we will
elaborate on the properties and phenomenological description of these
quantities, which can be regarded as crossed-channel analogs of GPDs.

The paper is organized as follows.  In the following two subsections
we define the kinematics and the observables for the process which we
will investigate, and recall its factorization property in Bjorken
kinematics.  In Section~\ref{sec_gpdmodel} we describe the model for
GPDs used in our work, and in Section~\ref{sec_2piDA} we give a
detailed discussion of two-pion distribution amplitudes and their
representation in terms of dispersion integrals.  The analytic form of
the scattering amplitude at NLO in $\alpha_s$ is given in
Section~\ref{sec_nlo}.  In Section~\ref{sec_2piModel} we develop a
number of model scenarios for $2\pi$DAs, and in
Section~\ref{sec_results} we present our results for observables in
two-pion production and their comparison with the HERMES data from
\cite{Airapetian:2004sy}.  We summarize our findings in
Section~\ref{sec_sum}.

%%%%%%%%%%%%%%%%%%%%%%%%%%%%%%%%%%%%%%%%%%%%%%%%%%%%%%%

\subsection{Kinematics and observables\label{sec_kin}}

\begin{figure}[t]
  \begin{center}
    \includegraphics[width=0.8\textwidth]{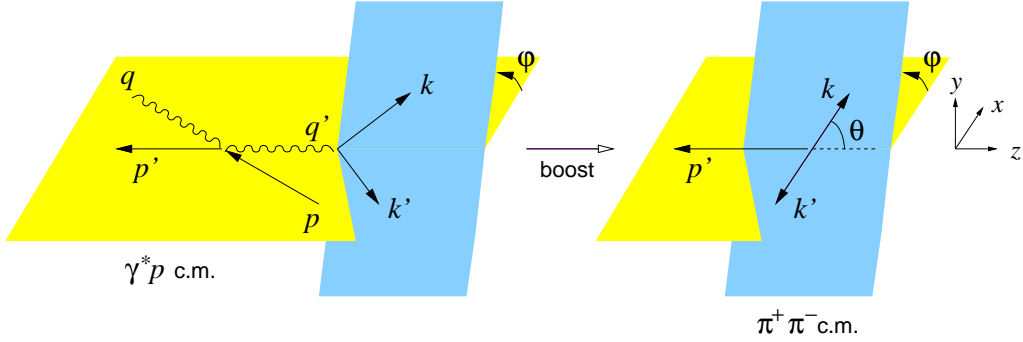}
    \caption{Definition of the angles $\theta$ and $\varphi$ in the
      two-pion system.
      \label{theta}
    }
  \end{center}
\end{figure}

We describe exclusive two-pion electroproduction on a nucleon using
the following momentum variables:
 \begin{equation}
  \label{prc}
   e(l)+N(p)\to e(l^\prime) +\pi^+(k)+\pi^-(k') + N(p^\prime) \,.
 \end{equation}
In the one-photon exchange approximation we can reduce our analysis to
the hadronic subprocess
 \begin{equation}
    \gamma^*(q) + N(p)\rightarrow \pi^+(k)+\pi^-(k') + N(p^\prime) \,.
    \label{subprc}
  \end{equation}
We specialize here to the case where the baryon is the same in the
initial and final state, but the theory description can be easily
extended to the case of a different outgoing baryon.  In addition, 
results for the production of two neutral pions can readily be
obtained from \eqref{prc} using isospin symmetry.
We use the conventional variables
 \begin{equation}
   \label{DISv}
      q= l-l^\prime \, , \quad q^2=-Q^2\, , \quad y=\frac{q \cdot p}
      {l\cdot p}\, , \quad
      W^2=(q+p)^2\, , \quad x_{B}=\frac{Q^2}{2p\cdot q}
 \end{equation}
for deep inelastic scattering, and in addition define the momentum
transfer 
\begin{equation}
  \label{tv}
  \Delta=p^\prime - p\, , \quad \Delta^2=t
\end{equation}
to the nucleon.  We denote the nucleon and pion mass by $m_N$ and
$m_\pi$, respectively, introduce the invariant mass of the two-pion
system,
\begin{equation}
  \label{mss}
  (k+k')^2= m_{\pi\pi}^2 =s_\pi^{} \,,
\end{equation}
and neglect the lepton mass throughout.  We also write $q' = k+k'$ for
the total momentum of the pion pair.  Finally, we define the polar and
azimuthal angles $\theta$ and $\varphi$ of the $\pi^+$ in the rest
frame of the two pions, as shown in Fig.~\ref{theta}.

An important set of observables for analyzing the two-pion system are
the Legendre moments
\begin{equation}
  \langle P_l(\cos \theta)\rangle =
    \frac{
      \int\limits_{-1}^1 \mathrm{d}\cos \theta\,\;
      P_l(\cos \theta)
      \dfrac{\mathrm{d} \sigma}{\mathrm{d} \cos \theta}
         }
         {
         \int\limits_{-1}^1 \mathrm{d}\cos \theta\,\;
         \dfrac{\mathrm{d} \sigma}{\mathrm{d} \cos \theta}
         } \,,
  \label{Pl}
\end{equation}
where $P_l$ is a Legendre polynomial.  To make their content explicit,
we decompose the cross section into partial waves of the produced
two-pion system,
\begin{align}
  \label{partial-waves}
\frac{1}{\sigma}\,
\frac{\diff\sigma}{\diff\cos\theta\; \diff\varphi}
&= \sum_{JJ^\prime\lambda\lambda^\prime}
  \rho_{\lambda\lambda^\prime}^{JJ^\prime}\,
Y_{J\lambda}^{}(\theta,\varphi)\,
Y_{J^\prime\lambda^\prime}^*(\theta,\varphi) \,,
&
\sum_{J \lambda} \rho_{\lambda\lambda}^{JJ} &= 1 \,,
\end{align}
where $\rho$ is the spin density matrix of the pion pair.  Its
diagonal elements $\rho_{\lambda\lambda}^{JJ}$ give the probability
that the two pions are in the state with total angular momentum $J$
and angular momentum component $\lambda$ along the $z$-axis in
Fig.~\ref{theta}, whereas the off-diagonal terms describe the
corresponding interference terms.  Neglecting $J>2$ contributions we
have in particular \cite{Sekulin:1973mk,Diehl:2003ny}
\begin{align}
  \label{Legendre-mom}
\langle P_1\rangle &=
\frac{1}{\sqrt{15}}\, \Re\left[4\sqrt{3}\,\rho_{11}^{21}
            +4\slim\rho_{00}^{21}+2\sqrt{5}\,\rho_{00}^{10}\right] \,,
\nonumber \\
\langle P_3\rangle &=
\frac{6}{7\sqrt{5}}\, \Re\left[-2\slim\rho_{11}^{21}
                   +\sqrt{3}\,\rho_{00}^{21}\right] \,.
\end{align}
Note that the combination
\begin{equation}
\big\langle P_1(\cos\theta)
        + \tfrac{7}{3}\slim P_3(\cos\theta) \big\rangle
= \frac{2}{\sqrt{3}}\,
  \Re \left[ \sqrt{5}\,\rho_{00}^{21} +\rho_{00}^{10} \right] 
\label{l_comb_0}
\end{equation}
projects out the state with zero total helicity of the final two
pions, whereas 
\begin{equation}
\big\langle P_1(\cos\theta) 
	- \tfrac{14}{9}\slim P_3(\cos\theta) \big\rangle 
= \frac{2}{3}\,
  \Re \left[ 2\sqrt{5}\,\rho_{11}^{21}
            +\sqrt{3}\,\rho_{00}^{10} \right]
\label{l_comb_1}
\end{equation}
involves the helicity-one but not the helicity-zero state of the two
pions in the $J=2$ partial wave.

%%%%%%%%%%%%%%%%%%%%%%%%%%%%%%%%%%%%%%%%%%%%%%%%%%%%%%%

\subsection{Factorization of the hadronic process\label{sec:fact}}

\begin{figure}
  \begin{center}
    \includegraphics[width=0.36\textwidth]{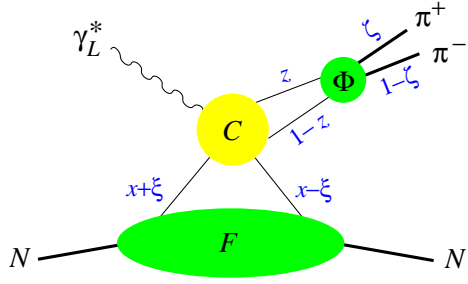}
    \caption{Factorization of hard exclusive pion pair production by a
      virtual photon.  $x$, $\xi$ and $z$, $\zeta$ are momentum
      fractions as explained in Sections~\protect\ref{sec_gpdmodel}
      and \protect\ref{sec_2piDA}. \label{fig_fact}}
  \end{center}
\end{figure}

We consider the hadronic process (\ref{subprc}) in the Bjorken limit,
where the scattering energy and the virtuality of the exchanged photon
are much larger than $\sqrt{-t}$, the nucleon mass, and the invariant
mass of the produced two-pion system,
\begin{equation}
    \label{Q2}
    W^2, Q^2\gg -t , m_N^2 , m^2_{\pi\pi} \,.
\end{equation}
According to the factorization theorem from
\cite{Collins:1996fb,Freund:1999xg} the scattering amplitude in this
limit can be written as the convolution of coefficient functions
$C^{ij}$ with nucleon matrix elements $F^i$ parameterized by GPDs and
with two-pion distribution amplitudes $\Phi^j$,
\begin{equation}
  \label{fact}
T = \frac{1}{Q}
\sum_{ij} \int_{-1}^1\mathrm{d}x \int_0^1 \mathrm{d}z \,
  F^i(x,\xi,t; \mu_F)\, C^{ij}(x,\xi,z; Q,\mu_F,\mu_R)\,
  \Phi^j(z,\zeta,s_\pi; \mu_F) + \mathcal{O}\left(\frac{1}{Q^2}\right) \,,
\end{equation}
where $i$ and $j$ stand for quarks or gluons, and $\mu_F$ and $\mu_R$
denote the factorization and renormalization scales.  More details on
$F^i$ and $\Phi^j$ and on their arguments $\xi$ and $\zeta$ will be
given in the following two sections.  The amplitude \eqref{fact}
refers to longitudinal polarization of the virtual photon and of the
pion pair, i.e.\ to $\lambda=0$ in \eqref{partial-waves}.  The
amplitudes involving a transverse photon or nonzero $\lambda$ decrease
at least like $1/Q^2$ and are hence power suppressed in Bjorken
kinematics.  The factorization formula \eqref{fact} is illustrated in
Fig.~\ref{fig_fact}.

We caution that the power corrections in \eqref{fact} need not be
numerically small for $Q^2$ of a few $\gev^2$.  Phenomenological
estimates based on transverse parton momentum effects in the hard
scattering \cite{Vanderhaeghen:1999xj,Goloskokov:2006hr} or on
renormalon calculations \cite{Vanttinen:1998pp,Belitsky:2003tm} have
indeed found substantial power corrections to the amplitude.  On the
other hand, the fair agreement of the leading-order calculation for
$\pi^+\pi^-$ production in \cite{Lehmann-Dronke:2000xq} with the
experimental results \cite{Airapetian:2004sy} for the Legendre moments
\eqref{Legendre-mom} gives hope that power corrections to these
observables may be not too large in the kinematics of the HERMES
measurement.

%%%%%%%%%%%%%%%%%%%%%%%%%%%%%%%%%%%%%%%%%%%%%%%%%%%%%%%

\section{Modeling the GPDs\label{sec_gpdmodel}}

Generalized parton distributions parameterize matrix elements of
light-cone separated quark or gluon operators.  The distributions
relevant for our process \eqref{subprc} are given by
\begin{align}
    F^q (x,\xi,t)
    &= \frac{1}{2} \int \frac{\diff\lambda}{2\pi}\,
    \e^{i x \lambda (P\cdot n)} \big\langle
    p^\prime \big|\, \bar q\bigl(-\half \lambda n\bigr) 
    \slim\slashed{n}\slim q \bigl(\half \lambda n\bigr) 
    \big| p\big\rangle
    \nonumber \\
    &= \frac{1}{2\slim P\cdot n} \left[
    H^q (x,\xi,t)\, \bar u(p^\prime) \slim\slashed{n}\slim u(p)
    + E^q (x,\xi,t)\, \bar u(p^\prime)
      \frac{i\sigma^{\alpha\beta}n_{\alpha}\Delta_\beta}{2 m_N} u(p)
    \right] \, ,
\nonumber \\
    F^g (x,\xi,t) 
    &= \frac{1}{P\cdot n} \int\frac{\diff\lambda}{2\pi}\,
    \e^{i x \lambda (P\cdot n)}\,
    n_{\alpha}n_{\beta}\,
    \big\langle
    p^\prime \big|G^{\alpha\mu} \bigl(-\half \lambda n\bigr)
    G_\mu{}^{\beta} \bigl(\half \lambda n\bigr)
    \big| p\big\rangle
    \nonumber \\
    &= \frac{1}{2\slim P\cdot n} \left[
    H^g (x,\xi,t)\, \bar u(p^\prime) \slim\slashed{n}\slim u(p)
    + E^g (x,\xi,t)\, \bar u(p^\prime) 
      \frac{i\sigma^{\alpha\beta}n_{\alpha}\Delta_\beta}{2 m_N} u(p)
    \right] \,,
    \label{GPD-def}
\end{align}
where $n$ is a lightlike auxiliary vector and we have omitted the
dependence on $\mu_F$ for simplicity.  For definiteness we consider
proton distributions in this section, the corresponding expressions
for a neutron target are readily obtained from isospin symmetry.  The
insertion of a Wilson line between the field operators is implied in
\eqref{GPD-def}.  The light-cone momentum fractions of the partons
with respect to the average nucleon momentum $P = \half (p+p')$ are
parameterized by $x$ and $\xi$ as shown in Fig.~\ref{fig_fact}.  The
skewness $\xi = - 2\slim (\Delta \cdot n)/(P \cdot n)$ is related to
the Bjorken variable in our process \eqref{subprc} by
\begin{equation}
\xi = \frac{x_B}{2- x_B} \,.
\end{equation}
In the forward limit, $p^\prime=p$, the distributions $E^q (x,\xi,t)$
and $E^g (x,\xi,t)$ decouple in the matrix elements \eqref{GPD-def},
whereas $H^q (x,\xi,t)$ and $H^g (x,\xi,t)$ reduce to the ordinary
parton densities
\begin{align}
H^q (x,0,0) &= q(x)       & \mbox{for}~ x>0 \, ,
\nonumber\\
H^q (x,0,0) &=-\bar q(-x) & \mbox{for}~ x<0 \, ,
\nonumber \\
H^g (x,0,0 )&= x g(x)     & \mbox{for}~ x>0 \, .
  \label{reduct}
\end{align}
Following the notation in \cite{Diehl:2003ny} we also use
combinations $F^{q(C)}$ of quark GPDs corresponding to $t$-channel
exchange with definite charge conjugation parity $C$,
\begin{align}
   \label{qC}
   F^{q(+)} (x,\xi,t)&=F^{q} (x,\xi,t)-F^{q} (-x,\xi,t) \, ,
\nonumber \\
   F^{q(-)} (x,\xi,t)&=F^{q} (x,\xi,t)+F^{q} (-x,\xi,t) \, ,
\end{align}
with analogous decompositions for the distributions $H$ and $E$.  The
gluon GPDs are even functions of $x$, i.e.\ $F^g(x,\xi,t) =
F^g(-x,\xi,t)$, and of course correspond to $C=+1$ exchange.

In our calculations we use the same model for the functions $H^q$ and
$H^g$ as in \cite{Lehmann-Dronke:2000xq}.  It is based on Radyushkin's
double distribution ansatz \cite{Radyushkin:1997ki,Musatov:1999xp},
supplemented by the Polyakov-Weiss term \cite{Polyakov:1999gs}.  For
the quark GPD at $t=0$ we write
\begin{equation}
    H^q(x,\xi,0)=H^q_\textrm{DD}(x,\xi,0)
    +\frac{1}{3}\slim \theta\bigl(\xi-\vert x\vert\bigr)\slim
    D\left(\frac{x}{\xi}\right)
    \label{modmodel}
\end{equation}
and use for the Polyakov-Weiss term the estimate \cite{Kivel:2000fg}
obtained in the chiral quark-soliton model,
\begin{equation}
    D(x)=-4.0\, (1-x^2)
    \left[
    C_1^{3/2}(x)+0.3\, C_3^{3/2}(x)+0.1\, C_5^{3/2}(x)
    \right] ,
    \label{Dterm}
\end{equation}
where $C_1^{3/2}(x)$ are Gegenbauer polynomials.  An analogous
representation holds for $H^g(x,\xi,0)$, where we set the
Polyakov-Weiss term to zero since there is no phenomenological
estimate available for it in the literature.  The double distribution
parts are written as
\begin{align}
  \label{dd-models}
H^q_\textrm{DD}(x,\xi,0)
 &= \int_{-1}^1 d\beta \int_{-1+|\beta|}^{1-|\beta|} d\alpha\;
  \delta(x-\beta-\xi\alpha)\,
  \Big[ \theta(\beta)\, q(\beta)\, 
      - \theta(-\beta)\, \bar{q}(-\beta) \Big]\, h(\beta,\alpha) ,
\nonumber \\
H^g_\textrm{DD}(x,\xi,0)
 &= \int_{-1}^1 d\beta \int_{-1+|\beta|}^{1-|\beta|} d\alpha\;
  \delta(x-\beta-\xi\alpha)\; \beta 
  \Big[ \theta(\beta)\, g(\beta)\, 
      - \theta(-\beta)\, g(-\beta) \Big]\, h(\beta,\alpha)
\end{align}
with a so-called profile function
\begin{equation}
  \label{profile}
h(\beta,\alpha) = \frac{\Gamma(2b+2)}{2^{2b+1}\slim [\Gamma(b+1)]^2}\,
\frac{[ (1-|\beta|)^2- \alpha^2 ]^{\slim b}}{(1-|\beta|)^{2b+1}} \,,
\end{equation}
where we take $b=1$ for quarks and $b=2$ for gluons.  For the ordinary
parton densities we will use the MRST 2004 NLO parameterization
\cite{Martin:2004ir} at scale $Q^2=3.2\, \gev^2$ or $Q^2=7\,\gev^2$,
depending on the value of $Q^2$ in the experimental observable to be
described.

For the $t$-dependence of the GPDs we use a factorized ansatz
\cite{Mankiewicz:1997aa,Vanderhaeghen:1998uc}.  This is known to be
oversimplified, both from general considerations
\cite{Burkardt:2004bv} and from lattice calculations
\cite{Gockeler:2005cd,Edwards:2006qx}.  In the present paper we will
however concentrate on observables such as the Legendre moments
\eqref{Legendre-mom}, where the bulk of the $t$-dependence in the GPDs
drops out and only details of this dependence matter that are
currently not well constrained by phenomenology.  For our purpose, a
factorized ansatz should hence be sufficient, and for the $C=-1$ quark
combination we write
\begin{equation}
  \label{dirac-ansatz}
H^{q(-)}(x,\xi,t) = H^{q(-)}(x,\xi,0)\, \frac{F_1^q(t)}{F_1^q(0)} \,,
\end{equation}
where $F_1^q(t)$ is the contribution from quark flavor $q$ to the
proton Dirac form factor.  This ansatz ensures the sum rule
\begin{equation}
  \label{F1-sum}
\int_{-1}^{1} \mathrm{d}x\, H^q(x,\xi,t)=F_1^q(t) \,.
\end{equation}
The input needed in \eqref{dirac-ansatz} can be obtained from the
measured electromagnetic nucleon form factors $F^p_{1} = \frac{2}{3}
F_1^u - \frac{1}{3} F_1^d - \frac{1}{3} F_1^s$ and $F^n_{1} =
\frac{2}{3} F_1^d - \frac{1}{3} F_1^u - \frac{1}{3} F_1^s$ combined
with lattice calculations of the strangeness form factors.  We have
taken the parameterizations from \cite{Ericson:1988my} and
\cite{Dong:1997xr}, which should be sufficiently reliable given the
limitations of the factorized form \eqref{dirac-ansatz}.
For the gluon GPD we make an analogous ansatz
\begin{equation}
H^{g}(x,\xi,t) = H^{g}(x,\xi,0)\, \frac{F_\theta(t)}{F_\theta(0)} \,,
\end{equation}
where $F_\theta(t)$ is a form factor of the gluon part of the
energy-momentum tensor, for which we use the model from
\cite{Braun:1992jp},
\begin{equation}
F_\theta (t) = F_\theta(0)\,
  \bigl[ 1-t/(3 M_\theta^2) \bigr]^{-3}
\end{equation}
with $M_\theta^2=2.6\mathrm{\,GeV^2}$.  The $C=+1$ quark combination
(which is not constrained by the Dirac form factors) is modeled by a
corresponding ansatz with the same form factor $F_\theta(t)$,
motivated by the fact that $H^{q(+)}(x,\xi,t)$ and $H^g(x,\xi,t)$ mix
under evolution.

We do not attempt to model $E^{q}$ and $E^g$ here and will neglect
their contribution when calculating observables.  This should be good
enough for our purposes, because for an unpolarized target they enter
with prefactors $t/(4 m_N^2)$ or $\xi^2$, which are small in the
kinematics we will consider.

%%%%%%%%%%%%%%%%%%%%%%%%%%%%%%%%%%%%%%%%%%%%%%%%%%%%%%%

\section{Two-pion distribution amplitudes\label{sec_2piDA}}

A central ingredient for the description of the process \eqref{prc}
are the two-pion distribution amplitudes.  For the $\pi^+\pi^-$ system
they are defined as \cite{Diehl:2003ny}
\begin{align}
\Phi^q(z,\zeta,s_\pi)
&= \int \frac{\diff \lambda}{2\pi}\,
\e^{-i z \lambda (q^\prime\cdot n)}\,
\big\langle
\pi^+(k)\slim\pi^-(k') \big|\bar q\bigl(\lambda n\bigr)
   \slashed{n}\slim q\bigl(0\bigr) \big|0\big\rangle \,,
\nonumber \\
\Phi^g(z,\zeta,s_\pi) &=
\frac{1}{q^\prime\cdot n} \int\frac{\diff\lambda}{2\pi}\,
\e^{-i z \lambda (q^\prime\cdot n)}\,
n_{\alpha}n_{\beta}\, \big\langle
\pi^+(k)\slim\pi^-(k') \big|G^{\alpha\mu} \bigl(\lambda n\bigr)
G_\mu{}^{\beta} \bigl(0\bigr)
\big| 0\big\rangle \,,
\label{GDA-def}
\end{align}
where $n$ is a lightlike auxiliary vector and we have suppressed the
dependence on the factorization scale $\mu_F$ as before.  As in the
case of GPDs, the insertion of an appropriate Wilson line between the
field operators is implied.  The $2\pi$DAs describe the exclusive
fragmentation of a pair of quarks or gluons into the final pion pair
\cite{Diehl:1998dk}.  The variable $z$ is the light-cone momentum
fraction of one of the two partons with respect to the total momentum
$q'$ of the pion pair.  The variable $\zeta$ characterizes the
distribution of the total momentum $q^\prime$ among the two pions,
\begin{equation}
    \zeta=\frac{k\cdot n}{q^\prime \cdot n} \,,
    \label{zeta}
\end{equation}
and is related to the polar angle $\theta$ and the relativistic
velocity of the $\pi^+$ in the c.m.\ of the pair by
\begin{align}
\label{beta-def}
  \beta \cos \theta &= 2\zeta-1 \,, &
  \beta &= \sqrt{1- \frac{4 m_\pi^2}{s_\pi}} \,.
\end{align}
It is useful to project out the combinations
\begin{equation}
\Phi^{q(\pm)}(z,\zeta,s_\pi) = \half \bigl[ 
  \Phi^{q}(z,\zeta,s_\pi) \pm \Phi^{q}(z,1-\zeta,s_\pi) \bigr]
\end{equation}
describing a two-pion system with definite charge conjugation parity
$C=\pm 1$.  Charge conjugation and isospin symmetry imply that
\begin{align}
  \label{C-da}
& \Phi^{I=0} = \Phi^{u(+)} = \Phi^{d(+)} \,,
&
& \Phi^{I=1} = \Phi^{u(-)} = -\Phi^{d(-)} \,,
\end{align}
where the combinations $\Phi^I$ associated with definite isospin $I$
of the pion pair have been introduced in \cite{Polyakov:1998ze}.  The
$2\pi$DAs for gluons and for strange quarks are of course pure
isosinglet.  We also remark that the distribution amplitude for $u$ or
for $d$ quarks in a $\pi^0\pi^0$ pair is equal to $\Phi^{I=0}$ by
isospin invariance.  

Following \cite{Polyakov:1998ze,Kivel:1999sd} we expand the
distribution amplitudes in Gegenbauer polynomials $C_n^m(2z-1)$ and
Legendre polynomials $P_l(2\zeta-1)$,
\begin{align}
\label{poly-expand}
    \Phi^{q(-)}(z,\zeta,s_\pi) &= 6z(1-z)
    \sum_{\substack{n=0\\\mathrm{even}}}^\infty
    \sum_{\substack{l=1\\\mathrm{odd}}}^{n+1}
    B_{nl}^{q(-)}(s_\pi)\, C_n^{3/2}(2z-1)\, P_l(2\zeta-1) \,,
\nonumber \\
    \Phi^{q(+)}(z,\zeta,s_\pi) &= 6z(1-z)
    \sum_{\substack{n=1\\\mathrm{odd}}}^\infty
    \sum_{\substack{l=0\\\mathrm{even}}}^{n+1}
    B_{nl}^{q(+)}(s_\pi)\, C_n^{3/2}(2z-1)\, P_l(2\zeta-1) \,,
\nonumber \\
    \Phi^g(z,\zeta,s_\pi) &= 9z^2(1-z)^2
    \sum_{\substack{n=1\\\mathrm{odd}}}^\infty
    \sum_{\substack{l=0\\\mathrm{even}}}^{n+1}
    B_{nl}^g(s_\pi)\, C_{n-1}^{5/2}(2z-1)\, P_l(2\zeta-1) \,,
\end{align}
where the restrictions to even or odd $n$ and $l$ follow from charge
conjugation invariance.\footnote{%
  The coefficients $B_{nl}^g$ follow the convention of
  \protect\cite{Diehl:2003ny} and are related to the coefficients
  $A_{nl}^G$ in \protect\cite{Kivel:1999sd} by $3 B_{nl}^g = 10
  A_{n-1, l}^G$.}
We will also use the notation
\begin{align}
& B^{I=0} = B^{u(+)} = B^{d(+)}\,,
&
& B^{I=1} = B^{u(-)} = -B^{d(-)}
\end{align}
corresponding to \eqref{C-da}.  The expansion of the $z$-dependence in
Gegenbauer polynomials is chosen such that to leading order in
$\alpha_s$ the coefficients $B_{nl}$ evolve multiplicatively in the
factorization scale $\mu_F$, with mixing occurring only between
$B_{nl}^g$ and the quark singlet combination $\sum_q
\smash{B_{nl}^{q(+)}}$, see e.g.\ \cite{Diehl:2000uv}.  The expansion
of the $\zeta$-dependence in Legendre polynomials is rather directly
related to the partial wave expansion of the two-pion system, as we
shall see shortly.

The coefficients $B_{nl}(s_\pi)$ parameterize matrix elements of local
operators between a $\pi^+\pi^-$ state and the vacuum, i.e.\ they are
form factors in the time-like region.  By analytic continuation they
are related to the spacelike form factors $A_{nk}(t_\pi)$ defined by
\begin{align}
  \label{GFFs}
  \big\langle \pi^+(p') \big| \bar{q}(0) \operatorname{\mathbf{S}}
     \gamma_{\mu_1}
     i\lrD_{\mu_2} \ldots i\lrD_{\mu_n}\slim q(0) 
     \big| \pi^+(p) \big\rangle
  &= 2 \sum_{\substack{k=0\\\mathrm{even}}}^n
        A_{nk}^q(t_\pi)\, \operatorname{\mathbf{S}}
           \Delta_{\mu_1} \ldots
           \Delta_{\mu_k} P_{\mu_{k+1}} \ldots P_{\mu_n} \,,
\nonumber \\
\big\langle \pi^+(p') \big| \operatorname{\mathbf{S}} G_{\mu_1\nu}(0)
    i\lrD_{\mu_2} \ldots i\lrD_{\mu_{n-1}} G^{\nu}{}_{\mu_n}(0)
      \big| \pi^+(p) \big\rangle
  &= 2 \sum_{\substack{k=0\\\mathrm{even}}}^n
        A_{nk}^g(t_\pi)\, \operatorname{\mathbf{S}}
           \Delta_{\mu_1} \ldots
           \Delta_{\mu_k} P_{\mu_{k+1}} \ldots P_{\mu_n} \,,
\end{align}
where $\lrD = \half \bigl( \rD - \lD \bigr)$, $P= \half(p+p')$,
$\Delta=p'-p$, $t_\pi=\Delta^2$, and $\operatorname{\mathbf{S}}$ denotes
symmetrization in all uncontracted Lorentz indices and subtraction of
trace terms.  These form factors are related to the Mellin moments of
pion GPDs.  For $t_\pi=0$ they reduce to the moments of the usual
quark and gluon densities in the pion, and one finds in particular
\begin{alignat}{2}
  \label{crossing}
  B^q_{n-1,n}(0) &= \frac{2}{3}\,\frac{2n+1}{n+1}\, A^q_{n0}(0)
  &&= \frac{2}{3}\,\frac{2n+1}{n+1}\int_0^1
    \diff x\, x^{n-1} \bigl[\slim
         q_\pi(x) + (-1)^n\slim \bar{q}_\pi(x) \slim\bigr] \,,
\nonumber \\
  B^g_{n-1,n}(0) &= \frac{8}{3}\,\frac{2n+1}{(n+1)(n+2)}\, A^g_{n0}(0)
  &&= \frac{8}{3}\,\frac{2n+1}{(n+1)(n+2)}
    \int_0^1 \diff x\, x^{n-1}\slim g_\pi(x) \,.
\end{alignat}
Phenomenological experience with the distribution amplitudes of single
mesons suggests that the coefficients of the expansion in Gegenbauer
polynomials decrease reasonably fast with $n$, see e.g.\
\cite{Ball:1996tb,Bakulev:2005cp,Braun:2006dg}.  This trend is
enhanced for larger factorization scales since the Gegenbauer
coefficients decrease faster with $\mu_F$ for increasing moment index
$n$.  Only the coefficient for $n=0$ and a linear combination of the
$n=1$ quark and gluon coefficients are independent of $\mu_F$ and
hence remain nonzero at asymptotically large $\mu_F$.  In our
phenomenological application we will only retain the $n=0$ and $n=1$
terms in \eqref{poly-expand}, keeping in mind that at moderately large
scales this may not be a very accurate approximation.  We then have
\begin{equation}
  \label{2piDA-I1}
  \Phi^{I=1}(z,\zeta,s_\pi)=6z(1-z)(2\zeta-1)\slim F_\pi(s_\pi) \,,
\end{equation}
where we have identified $B_{01}^{I=1}(s_\pi)$ with the
electromagnetic pion form factor $F_\pi(s_\pi)$, and
\begin{align}
  \label{2piDA-I0}
\Phi^{I=0}(z,\zeta,s_\pi) &= 18 z(1-z) (2z-1) 
  \Bigl[ B_{10}^{I=0}(s_\pi)
       + B_{12}^{I=0}(s_\pi)\, P_2(2\zeta-1) \Bigr] \,,
\nonumber \\[0.3em]
\Phi^g(z,\zeta,s_\pi) &= 9 z^2(1-z)^2\slim
  \Bigl[ B^g_{10}(s_\pi)
       + B^g_{12}(s_\pi)\, P_2(2\zeta-1) \Bigr] \,.
\end{align}
Notice that $F_\pi(s_\pi)$ is independent of the factorization scale
since it is associated with the conserved quark vector current.  The
coefficients $B_{1l}^{q}$ and $B_{1l}^g$ depend on $\mu_F$ in the same
way as the quark and gluon momentum fractions $\int_0^1 \diff x\, x
\bigl[ q_\pi(x) + \bar{q}_\pi(x) \bigr]$ and $\int_0^1 \diff x\, x
g_\pi(x)$, in accordance with \eqref{crossing}.  The sum $B^g_{1l} +
\sum_q B^q_{1l}$ is again $\mu_F$ independent since it is associated
with the \emph{total} energy-momentum tensor.  The coefficients in
\eqref{2piDA-I0} are related to the form factors in \eqref{GFFs}
by\,\footnote{%
  We remark that the coefficients in eq.~(90) of
  \protect\cite{Diehl:2003ny} should read $\frac{10}{9}$ and not
  $\frac{9}{10}$.}
\begin{align}
  \label{BtoA}
B_{10}(s_\pi) &= \tfrac{5}{9} A_{20}(s_\pi) 
               + \tfrac{20}{3} A_{22}(s_\pi) \,,
&
B_{12}(s_\pi) &= \tfrac{10}{9} A_{20}(s_\pi)
\end{align}
for both quarks and gluons.  Chiral dynamics constrains these form
factors for $|s_\pi| \ll \Lambda_\chi$, where $\Lambda_\chi = 4\pi
f_\pi \approx 1.16 \gev$ is the characteristic scale of chiral
symmetry breaking
\cite{Polyakov:1998ze,Kivel:2002ia,Lehmann-Dronke:2000xq}.  {}From the
one-loop calculation \cite{Diehl:2005rn} in chiral perturbation theory
we obtain
\begin{align}
  \label{chpt}
B_{10}^{}(s_\pi) &= - B_{12}^{(0)}\,
  \biggl\{ 1 + c_{10}^{(m)} m_\pi^2 + c_{10}^{(s)}\slim s_\pi^{}
   + \frac{m_\pi^2 - 2 s_\pi^{}}{2 \Lambda_\chi^2} \left[ 
       \ln\frac{m_\pi^2}{\mu_\chi^2}
     + \frac{4}{3}
     - \frac{s_\pi^{} + 2 m_\pi^2}{s_\pi}\, J(\beta) \right]
  \biggr\} + O\bigl( \Lambda_\chi^{-4} \bigr) \,,
\nonumber \\[0.3em]
B_{12}^{}(s_\pi) &= B_{12}^{(0)}\, \biggl\{ 1 
  + c_{12}^{(m)} m_\pi^2 + c_{12}^{(s)}\slim s_\pi^{} \biggr\}
  + O\bigl( \Lambda_\chi^{-4} \bigr) \,,
\end{align}
where $J(\beta) = 2 + \beta \ln\bigl[ (\beta-1) /(\beta+1) \bigr]$
with $\beta$ defined in \eqref{beta-def}, and $c_{1l}^{(m)}$,
$c_{1l}^{(s)}$ are unknown low-energy constants, whose natural size is
$\Lambda_\chi^{-2}$.  We have not displayed the dependence of
$\smash{c_{10}^{(m)}}$ and $\smash{c_{10}^{(s)}}$ on the
renormalization scale $\mu_\chi$, which cancels against the explicit
logarithm in \eqref{chpt}.  We recover the soft-pion theorem
$B_{10}(0) \approx - B_{12}(0)$ from \cite{Polyakov:1998ze} and obtain
the leading chiral correction to it:
\begin{align}
B_{10}(0) &= - B_{12}(0)\,
  \left\{ 1 + \frac{m_\pi^2}{2\Lambda_\chi^2}
              \left[ \ln\frac{m_\pi^2}{\mu_\chi^2} + 1 \right]
          +  m_\pi^2\slim \Bigl[ c_{10}^{(m)} - c_{12}^{(m)} \Bigr]
  \right\} + O\bigl( \Lambda_\chi^{-4} \bigr) \,.
\end{align}

%%%%%%%%%%%%%%%%%%%%%%%%%%%%%%%%%%

\subsection{Partial wave decomposition and Omn\`es
  representation\label{Omnes}}

The expansion \eqref{poly-expand} of the $\zeta$ dependence in
Legendre polynomials resembles a partial wave decomposition of the
two-pion system, which expands in $P_l(\cos\theta)$.  Indeed one can
readily rewrite the polynomials $P_l(2\zeta-1) = P_l(\beta
\cos\theta)$ in terms of $P_{k}(\cos\theta)$ with $k\le l$.  For $n=2$
one obtains \cite{Diehl:2000uv}
\begin{equation}
  \label{zeta-beta}
B_{10}(s_{\pi}) + B_{12}(s_{\pi})\, P_2(2\zeta-1)
= \tilde{B}_{10}(s_{\pi}) + \tilde{B}_{12}(s_{\pi})\, P_2(\cos\theta) \,,
\end{equation}
where the new coefficients
\begin{align}
  \label{Btilde}
\tilde{B}_{10}(s_{\pi}) 
&= {B}_{10}(s_{\pi}) - \frac{1-\beta^2}{2}\, {B}_{12}(s_{\pi})
 = {B}_{10}(s_{\pi}) - \frac{2m_\pi^2}{s_{\pi}}\slim {B}_{12}(s_{\pi}) ,
&
\tilde{B}_{12}(s_{\pi}) &= \beta^2\slim {B}_{12}(s_{\pi})
\end{align}
describe the two pions in an $S$ and a $D$ wave, respectively.  This
holds for both quark and gluon coefficients, and we drop the
corresponding superscript in the present subsection.  The phase of
$\tilde{B}_{nl}(s_{\pi})$ reflects the interaction of two pions in the
partial wave $l$.  For values of $s_{\pi}$ where $\pi\pi$ scattering is
elastic, one can apply Watson's theorem and finds
\cite{Polyakov:1998ze}
\begin{equation}
  \label{Watson}
\tilde{B}_{nl}^*(s_{\pi}) 
  = \tilde{B}_{nl}^{}(s_{\pi})\slim \exp\bigl[-2i \delta_l(s_{\pi})\bigr] \,,
\end{equation}
where $\delta_l(s_{\pi})$ is the phase shift for elastic $\pi\pi$ scattering
in the appropriate isospin channel ($I=0$ for even $l$ and $I=1$ for
odd $l$).  This relation determines the phase of $\tilde{B}_{nl}$ up
to a multiple of $\pi$.

The form factors $B_{nl}(s_{\pi})$ satisfy the usual analyticity properties
in $s_{\pi}$, i.e.\ they have a branch cut on the real axis above threshold
($s_{\pi}\ge 4 m_\pi^2$) and are real-valued for real $s_{\pi}$ 
below threshold.
Together with the phase information from Watson's theorem, one can
write down an Omn\`es representation for the form factors, as was
first pointed out in \cite{Polyakov:1998ze}.  We need to review this
issue here and start with a derivation of the Omn\`es representation
in a form adapted to our purpose.  Let $F(s_{\pi})$ be a form factor which
is nonzero at $s_{\pi}=0$ and at $s_{\pi}=4 m_\pi^2$ and has the following
properties:
\begin{enumerate}
\item $F(s_{\pi})$ is analytic in the $s_{\pi}$ plane except for a cut
  along the 
  real axis for $s_{\pi}\ge 4m_\pi^2$, and it is real-valued for real 
  $s_{\pi} <4 m_\pi^2$.
\item The complex phase of $F(s_{\pi}) /F(4 m_\pi^2)$ is $\delta_F(s_{\pi})$.
  With Watson's theorem we will have $\delta_F(s_{\pi}) = 
\delta_l(s_{\pi})$ for
  $s_{\pi}$ in the region above threshold where $\pi\pi$ scattering is
  elastic.
\item $\delta_F(s_{\pi})$ tends to a constant for $|s_{\pi}|\to \infty$.
\item $F(s_{\pi})$ has a finite number of simple zeroes at 
$s_{\pi}=s_1$, $s_{\pi}=s_2$,
  \ldots, $s_{\pi}=s_n$, where $n$ may also be zero.  Because of property 1
  the $s_i$ are either real-valued or come in complex conjugate pairs
  $s_{i+1}^* = s_i^{}$.
\end{enumerate}
We now consider
\begin{equation}
  \label{log-ff}
G(s_{\pi}) = \ln \frac{F(s_{\pi})}{F(0)\,
 (1-s_{\pi}/s_1) (1-s_{\pi}/s_2) \ldots (1-s_{\pi}/s_n)} ,
\end{equation}
where the Riemann sheet of the complex logarithm is chosen such that
$G(s_{\pi})$ is continuous and that $G(0) = 0$.  Then $G(s_{\pi})$ has
the same 
analyticity properties as $F(s_{\pi})$.  Note that for this it was necessary
to divide out possible zeroes of $F(s_{\pi})$ before taking the logarithm.
One can now write down a dispersion relation with $N \ge 1$
subtractions,
\begin{equation}
  G(s_{\pi}) = 
  \sum_{k=1}^{N-1} \frac{s_{\pi}^k}{k!}\, 
  \frac{\diff^k}{\diff s_{\pi}^k}\, G(0)
  + \frac{s_{\pi}^N}{\pi} \int_{4m_\pi^2}^\infty \diff s\, 
  \frac{\delta_F(s)}{{s}^{N} (s-s_{\pi}-i\varepsilon)} \,,
\end{equation}
where we have used that $\Im G(s_{\pi}) = \delta_F(s_{\pi})$ for real
$s_{\pi} \ge 4 
m_\pi^2$.  A term with $k=0$ does not appear in the sum because
$G(0)=0$.  The term $i\varepsilon$ implements the usual prescription
for handling the singularity at $s=s_{\pi}$.  We thus have
\begin{align}
  \label{new-omnes}
F(s_{\pi}) &= F(0)\, (1-s_{\pi}/s_1) (1-s_{\pi}/s_2) \ldots (1-s_{\pi}/s_n)
\nonumber \\[0.2em]
 & \quad\times \exp\Bigg[
   \sum_{k=1}^{N-1} \frac{s_{\pi}^k}{k!}\, 
   \frac{\diff^k}{\diff s_{\pi}^k}\, G(0)
 + \frac{s_{\pi}^N}{\pi} \int_{4m_\pi^2}^\infty \diff s\,
   \frac{\delta_F(s)}{{s}^{N} (s-s_{\pi}-i\varepsilon)} \Bigg] .
\end{align}
This representation can readily be used for $F(s_{\pi}) = B_{12}(s_{\pi})$.
Assuming that $B_{12}(s_{\pi})$ has no zero, we recover the representation
of this form factor already given in \cite{Polyakov:1998ze},
\begin{equation}
  \label{omnes-d}
\tilde{B}_{12}(s_{\pi}) = \beta^2 {B}_{12}(0)\slim f_2(s_{\pi})
\end{equation}
with the Omn\`es function
\begin{align}
  \label{omnes-fct-d}
f_2(s_{\pi}) &= \exp\Biggl[ \frac{s_{\pi}}{\pi} 
          \int_{4m_\pi^2}^\infty \diff s\, 
            \frac{\tilde{\delta}_2(s)}{{s} (s-s_{\pi}-i\varepsilon)}
          \Biggr]
\nonumber \\
&= \exp\Biggl[ s_{\pi}\, \frac{\diff}{\diff s_{\pi}} \ln B_{12}(0)
  + \frac{s_{\pi}^2}{\pi} \int_{4m_\pi^2}^\infty \diff s\, 
    \frac{\tilde{\delta}_2(s)}{{s}^{2} (s-s_{\pi}-i\varepsilon)} 
  \Biggr] \,,
\end{align}
where $\tilde{\delta}_2(s_{\pi})$ is the phase of ${B}_{12}(s_{\pi})
/{B}_{12}(4 m_\pi^2)$ and the forms with $N=1$ and $N=2$
subtractions are simultaneously valid.  At small enough $s_{\pi}$ one has
$\tilde{\delta}_2(s_{\pi}) = \delta_{2}(s_{\pi})$.

For the $S$ wave the situation is more involved.  To make use of the
phase information from Watson's theorem one has to consider
$\tilde{B}_{10}(s_{\pi})$, which has a pole at $s_{\pi}=0$ according to
\eqref{Btilde}.  We can however use the representation
\eqref{new-omnes} for
\begin{equation}
F(s_{\pi}) = s_{\pi}\tilde{B}_{10}(s_{\pi}) = s_{\pi} B_{10}(s_{\pi}) - 
2m_\pi^2 B_{12}(s_{\pi}) \,.
\end{equation}
According to the result \eqref{chpt} from chiral perturbation theory,
this form factor has a zero for $s_1 \approx - 2m_\pi^2$.  For the
following it is convenient to write
\begin{equation}
  \label{Fzero}
s_1 = \frac{2m_\pi^2}{1+\epsilon}\, \frac{B_{12}(0)}{B_{10}(0)} \,,
\end{equation}
so that $F(0)\, (1-s_{\pi}/s_1) = - 2m_\pi^2 B_{12}(0) + (1+\epsilon)\slim
s_{\pi}\slim B_{10}(0)$.  If we assume that $s_{\pi}
\tilde{B}_{10}(s_{\pi})$ only has 
the zero just discussed, we have
\begin{equation}
  \label{log-deriv}
\frac{\diff}{\diff s_{\pi}}\, G(0) = 
\frac{\diff}{\diff s_{\pi}} \ln B_{12}(0) 
  + \frac{\epsilon}{2m_\pi^2}\, \frac{B_{10}(0)}{B_{12}(0)} \,,
\end{equation}
where both terms are of order $\Lambda_\chi^{-2}$.  The Omn\`es
representation (\ref{new-omnes}) for $F(s_{\pi}) =
s_{\pi}\tilde{B}_{10}(s_{\pi})$ now gives
\begin{equation}
  \label{omnes-s}
\tilde{B}_{10}(s_{\pi}) =
  - {B}_{12}(0)\, \frac{3 C - \beta^2}{2}\slim f_0(s_{\pi})
\end{equation}
with
\begin{align}
  \label{C-def}
C &= \frac{1}{3} 
   - \frac{2 (1+\epsilon)}{3}\, \frac{{B}_{10}(0)}{B_{12}(0)}
\\
\intertext{and}
  \label{omnes-fct-s}
f_0(s_{\pi}) &= \exp\Biggl[ \frac{s_{\pi}}{\pi}
          \int_{4m_\pi^2}^\infty \diff s\, 
            \frac{\tilde{\delta}_0(s)}{{s} (s-s_{\pi}-i\varepsilon)}
          \Biggr]
\nonumber \\
&= \exp\Biggl[ s_{\pi}\, \bigg\{ \frac{\diff}{\diff s_{\pi}} \ln B_{12}(0) 
  + \frac{\epsilon}{2m_\pi^2}\, \frac{B_{10}(0)}{B_{12}(0)} \slim\bigg\}
  + \frac{s_{\pi}^2}{\pi} \int_{4m_\pi^2}^\infty \diff s\, 
    \frac{\tilde{\delta}_0(s)}{{s}^{2} (s-s_{\pi}-i\varepsilon)} 
  \Biggr] \,,
\end{align}
where $\tilde{\delta}_0(s_{\pi})$ is the phase of $\tilde{B}_{10}(s_{\pi})
/\tilde{B}_{10}(4 m_\pi^2)$.  With \eqref{chpt} the constants
appearing in the Omn\`es representation can be expressed as
\begin{align}
  \label{G-chiral}
\frac{\diff}{\diff s_{\pi}}\, G(0) &=
  - \frac{1}{\Lambda_\chi^2} \left[
    \ln\frac{m_\pi^2}{\mu_\chi^2}
    + \frac{17}{12} \right]
  + c_{10}^{(s)} + O\bigl( \Lambda_\chi^{-4} \bigr) \,,
\\[0.2em]
  \label{C-chiral}
C &=
  1 + \frac{5 m_\pi^2}{3 \Lambda_\chi^2} \left[
      \ln\frac{m_\pi^2}{\mu_\chi^2} + \frac{4}{3} \right]
  + \frac{2 m_\pi^2}{3}\, \Bigl[ c_{10}^{(m)} - c_{12}^{(m)}
           - 2 c_{10}^{(s)} + 2 c_{12}^{(s)} \Bigr]
  + O\bigl( \Lambda_\chi^{-4} \bigr) \,.
\end{align}
If we take $\mu_\chi = m_\rho$ then the first term in \eqref{G-chiral}
gives $-1.5 \gev^{-2}$ and the first two terms in \eqref{C-chiral}
give $0.95$.  Without further dynamical input we can of course not
estimate the values of the low-energy constants.

Inserting \eqref{omnes-d} and \eqref{omnes-s} into \eqref{zeta-beta},
we obtain
\begin{equation}
  \label{omnes-total}
B_{10}(s_{\pi}) + B_{12}(s_{\pi})\, P_2(2\zeta-1)
= - {B}_{12}(0) \left[ \frac{3 C - \beta^2}{2}\slim f_0(s_{\pi})
    - \beta^2\slim f_2(s_{\pi})\, P_2(\cos\theta) \right]
\end{equation}
with the Omn\`es functions $f_0$ and $f_2$ given in
\eqref{omnes-fct-s} and \eqref{omnes-fct-d} and $C$ specified by
\eqref{C-def} and \eqref{Fzero}.  This coincides with the
representation given without derivation in \cite{Kivel:1999sd} and
used in \cite{Lehmann-Dronke:1999aq,Lehmann-Dronke:2000xq}.  The paper
\cite{Kivel:1999sd} did not explicitly define the coefficient $C$, but
for the quark isosinglet case it quoted an estimate $C = 1 - b m_\pi^2
+ O(m_\pi^4)$ with $b \approx -1.7 \gev^{-2}$ from an instanton model
calculation, which results in $C \approx 0.97$.  The Omn\`es functions
used in
\cite{Kivel:1999sd,Lehmann-Dronke:1999aq,Lehmann-Dronke:2000xq} were
for $N=1$ subtraction and coincide with the first lines of
\eqref{omnes-fct-s} and \eqref{omnes-fct-d}.  For $N>1$ subtractions
our result differs from the dispersion relation for $B_{10}(s_{\pi})$ given
in these papers, where the transformation from $B_{10}(s_{\pi})$ to
$\tilde{B}_{10}(s_{\pi})$ and the zero of $\tilde{B}_{10}(s_{\pi})$ at
$s_{\pi} \approx 
-2m_\pi^2$ is not discussed.  In the Omn\`es function $f_0(s_{\pi})$ this
would result in a term $s_{\pi} \frac{\diff}{\diff s_{\pi}} \ln B_{10}(0)$
instead of $s_{\pi} \smash{\frac{\diff}{\diff s_{\pi}}} G(0)$ specified by
\eqref{log-deriv}.  Using the result \eqref{chpt} from chiral
perturbation theory we find that
\begin{equation}
  \label{B12-10}
\frac{\diff}{\diff s_{\pi}} \ln B_{10}(0)
= \frac{\diff}{\diff s_{\pi}}\, G(0) 
  + \frac{19}{60}\slim \frac{1}{\Lambda_\chi^2}
  + O\bigl( \Lambda_\chi^{-4} \bigr)
\end{equation}
with $\smash{\frac{\diff}{\diff s}} G(0)$ given in
\eqref{G-chiral}.  Numerically, the second term on the r.h.s.\ is
$0.23 \gev^{-2}$.

The results of this subsection can be applied to both the isosinglet
quark and to the gluon form factors, where the forward limits
\begin{align}
  \label{forwardB}
B_{12}^{I=0}(0) &= \tfrac{10}{9} A_{20}^u(0) 
                 = \tfrac{10}{9} A_{20}^d(0)
&
B_{12}^g(0) &= \tfrac{10}{9} A_{20}^g(0)
\\
\intertext{with}
A_{20}^q(0) &= \int_0^1 \diff x\, x
  \bigl[ q_\pi(x) + \bar{q}_\pi(x) \bigr] \,,
&
A_{20}^g(0) &= \int_0^1 \diff x\, x g_\pi(x)
\end{align}
are of order one.  In contrast, $B_{12}^s(0)$ is $10/9$ times the
momentum fraction carried by strange quarks and antiquarks in a pion
and therefore quite small.  This calls for a careful analysis of the
size of different terms in the chiral expansion of
$B_{12}^s(s_\pi)$.  We shall not pursue this issue here since we will
not include the $2\pi$DA for strangeness in our phenomenological
application.

%%%%%%%%%%%%%%%%%%%%%%%%%%%%%%%%%%%%%%%%%%%%%%%%%%%%%%%

\section{The scattering amplitude at NLO\label{sec_nlo}}

In this section we give the expression of the scattering amplitude for
two-pion production to leading power in $1/Q$ and to NLO in
$\alpha_s$.  We decompose the amplitude for $\gamma^* N\to \pi^+\pi^-
N$ into terms $T^{(C)}$ describing a two-pion state with definite
charge conjugation parity $C=-1$ or $C=+1$,
\begin{equation}
    T = T^{(-)} + T^{(+)} \,.
\end{equation}
We note that the corresponding amplitude for $\gamma^* N\to \pi^0\pi^0
N$ is simply given by $T^{(+)}$.  To leading power in $1/Q$ we have
\begin{align}
  \label{I1}
  T^{(-)} =
  \frac{2\pi\sqrt{4\pi\alpha}}{N_c\slim \xi Q} \int^1_{-1} \diff x
&  \int^1_{0} \diff z
  \sum_{q=u,d} e_q \Phi^{q(-)}(z,\zeta,s_\pi) \,
  \Biggl[ Q^{(+)}(z,x/\xi)\, F^{q(+)} (x,\xi,t)
\nonumber \\
& + G^{(+)}(z,x/\xi)\, \frac{1}{2\xi}\, F^{g}(x,\xi,t)
  + R^{(+)}(z,x/\xi)
    \sum_{q^\prime=u,d,s} F^{q^\prime (+)} (x,\xi,t)
  \Biggr] ,
\\
\intertext{and}
  \label{I0}
  T^{(+)} = 
  \frac{2\pi\sqrt{4\pi\alpha}}{N_c\slim \xi Q} \int^1_{-1} \diff x
&  \int^1_{0} \diff z
  \sum_{q=u,d,s} e_q F^{q(-)}(x,\xi,t) \,
  \Biggl[ Q^{(-)}(z,x/\xi)\, \Phi^{q(+)}(z,\zeta,s_\pi)
\nonumber \\
& + G^{(-)}(z,x/\xi)\, \Phi^g(z,\zeta,s_\pi)
  + R^{(-)}(z,x/\xi) 
    \sum_{q^\prime=u,d,s} \Phi^{q^\prime (+)}(z,\zeta,s_\pi)
  \Biggr] ,
\end{align}
where $N_c=3$ denotes the number of colors, $\alpha$ the fine
structure constant, and $e_q$ the quark charge in units of the
positron charge.  As discussed in Section~\ref{sec:fact}, these
leading amplitudes in $1/Q$ are for longitudinal photon polarization.

\begin{figure}
\begin{center}
\subfigure[]
{\includegraphics[width=0.43\textwidth,clip]{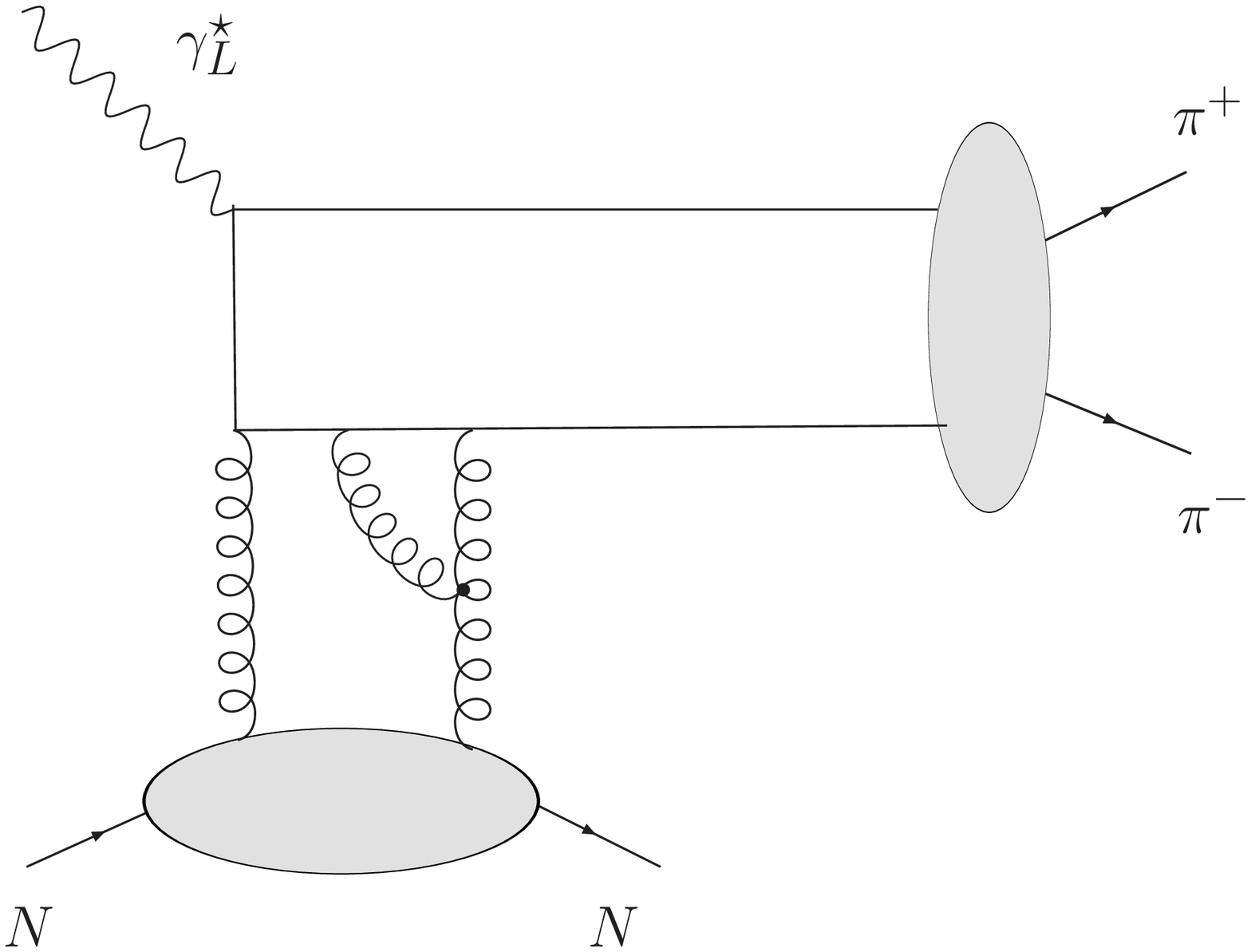}
\label{I1_a}}
\subfigure[]
{\includegraphics[width=0.43\textwidth,clip]{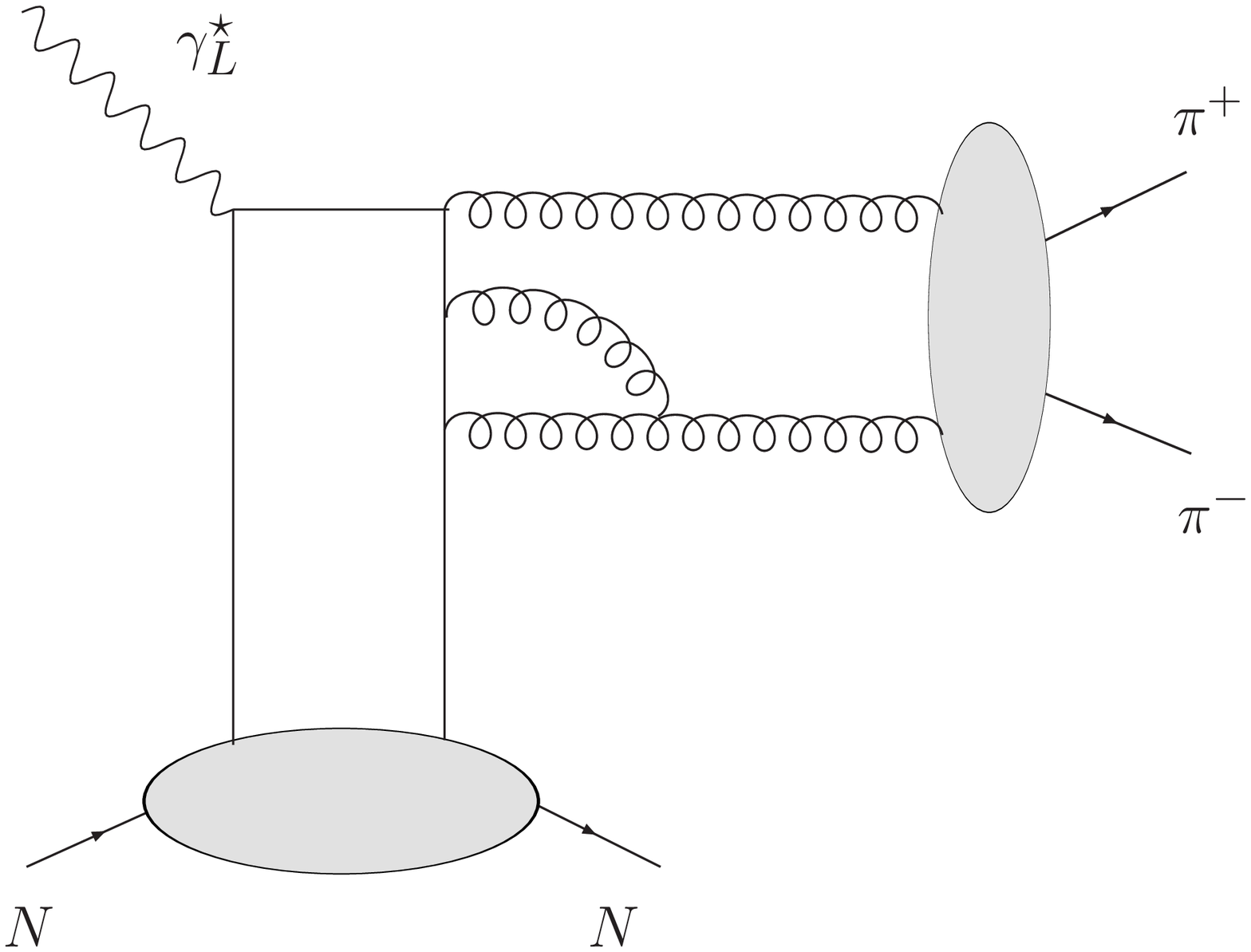}
\label{I0_a}}\\
\subfigure[]
{\includegraphics[width=0.43\textwidth,clip]{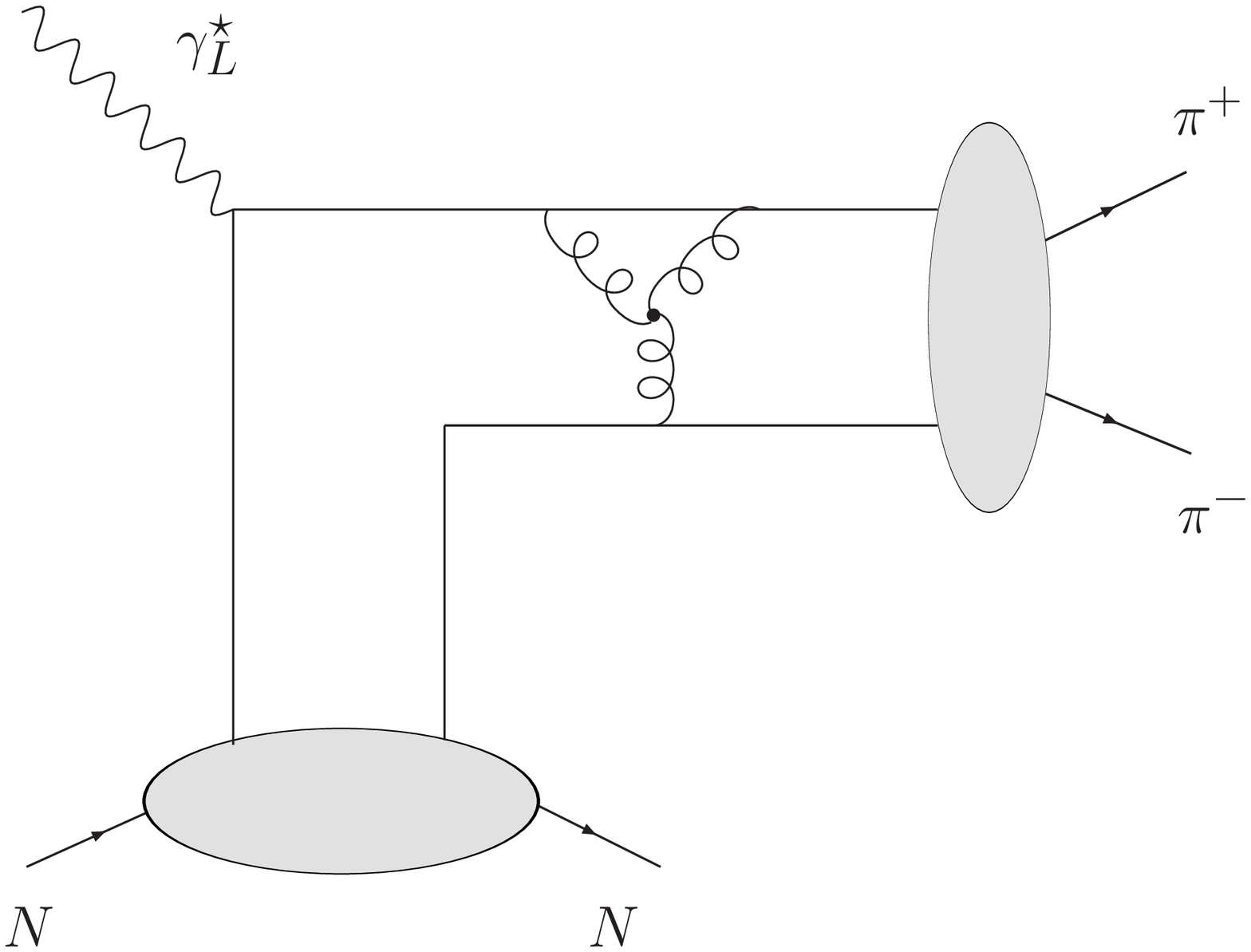}
\label{I1_b}}
\subfigure[]
{\includegraphics[width=0.43\textwidth,clip]{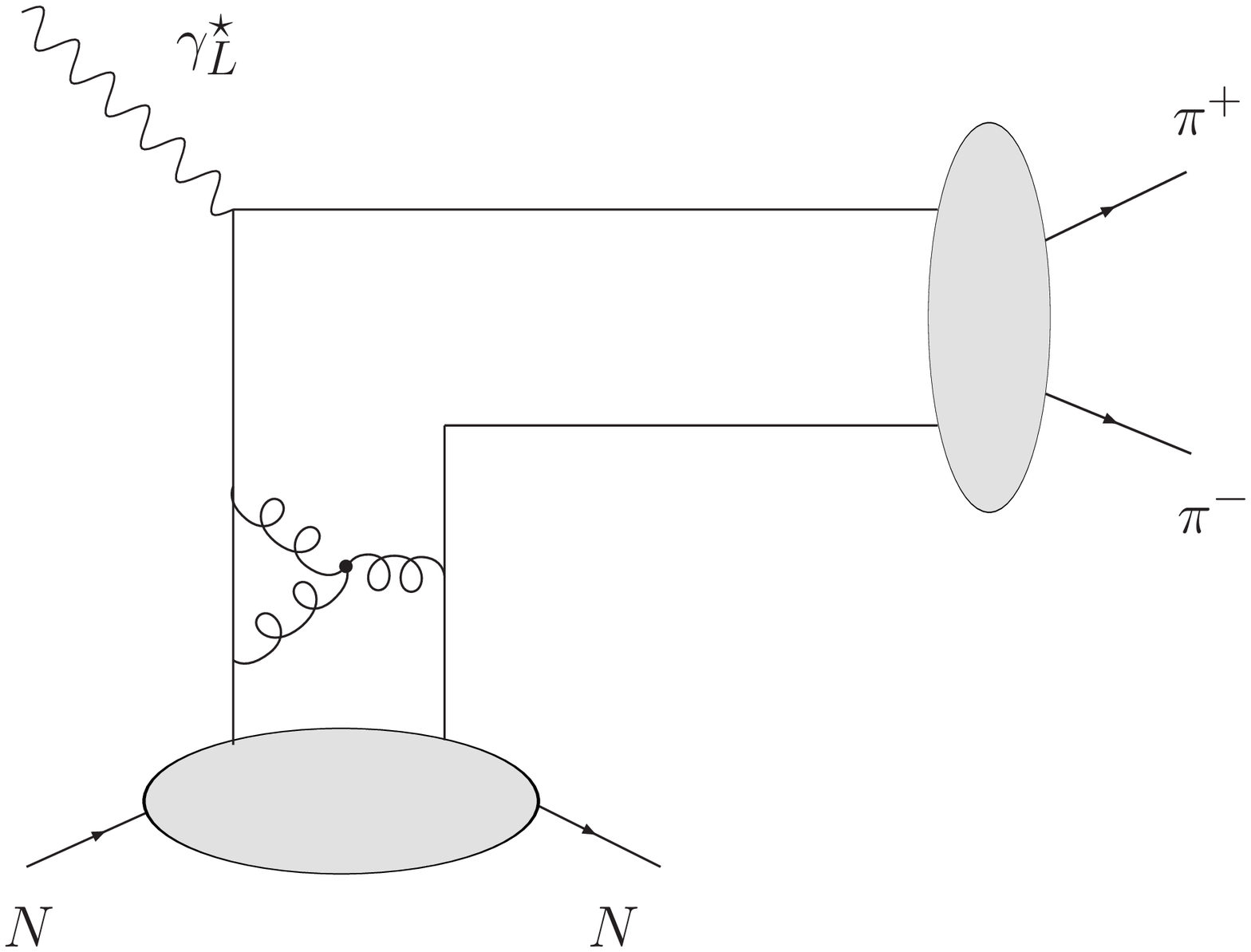}
\label{I0_b}}\\
\subfigure[]
{\includegraphics[width=0.43\textwidth,clip]{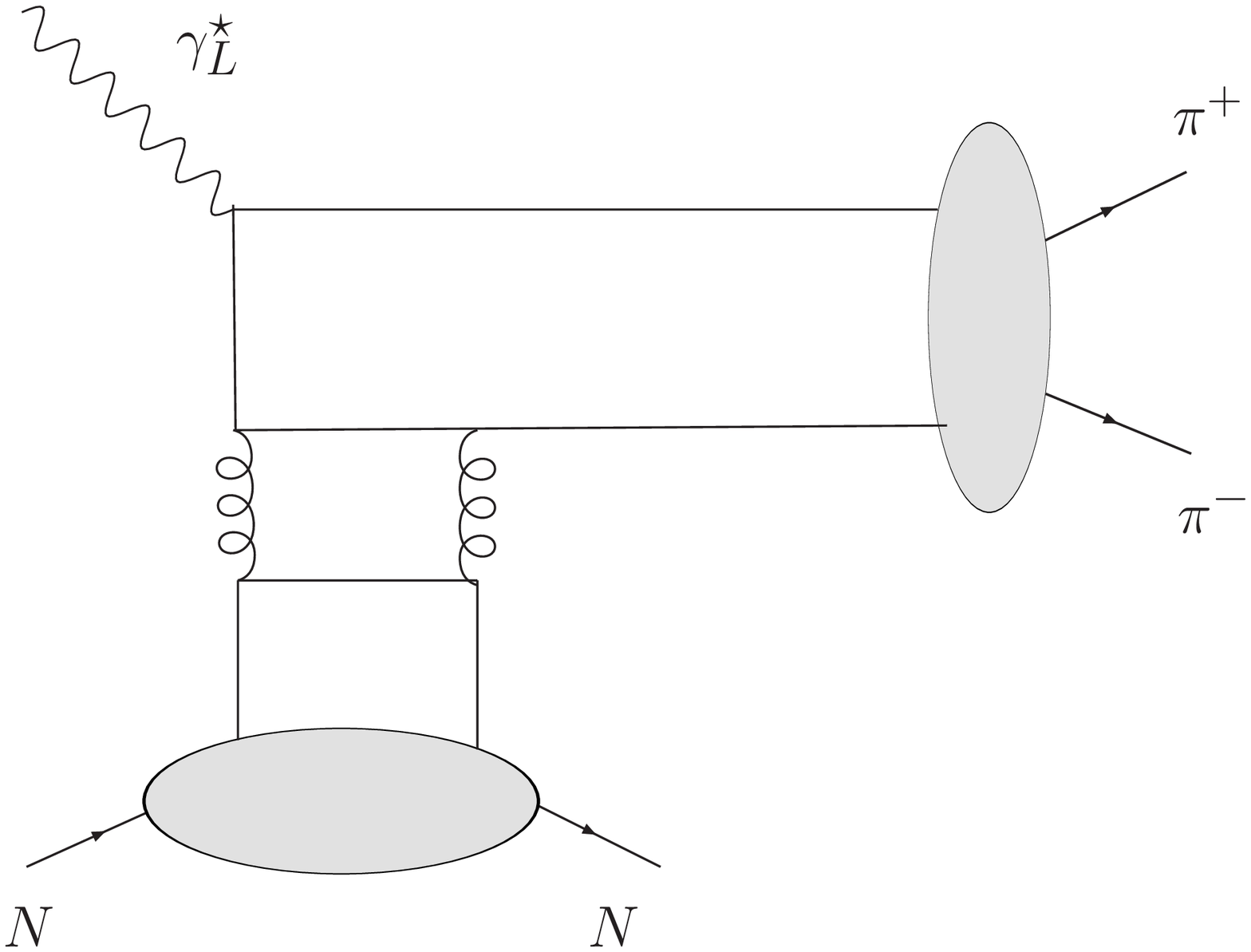}
\label{I1_c}}
\subfigure[]
{\includegraphics[width=0.43\textwidth,clip]{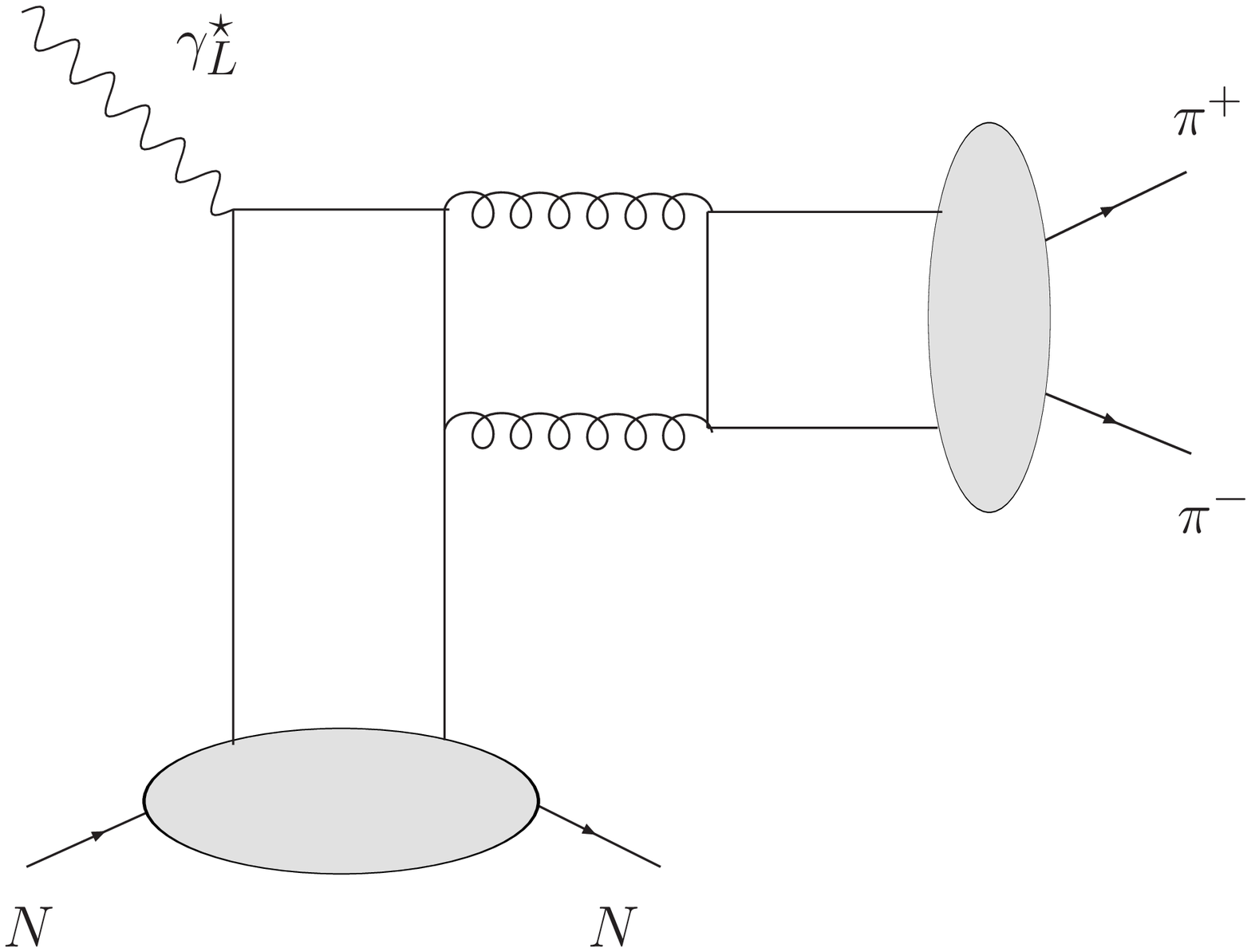}
\label{I0_c}}
\caption{\label{fig_I} Typical NLO diagrams for the production of two
  pions in a state with $C=-1$ (left side) or $C=+1$ (right side).
  Diagrams related to each other by crossing are displayed side by
  side.}
\end{center}
\end{figure}

The coefficient functions $R^{(\pm)}, G^{(\pm)}$ and $Q^{(\pm)}$
represent the amplitudes for the scattering of collinear partons, with
the appropriate subtraction of ultraviolet and collinear singularities
performed in the $\overline{\textrm{MS}}$ scheme.  $R^{(+)}$ and
$G^{(+)}$ were calculated in \cite{Ivanov:2004zv}, where
electroproduction of light vector mesons was studied at NLO.  The
coefficient function $Q^{(+)}$ for singlet quark exchange in the
$t$-channel (a typical diagram is shown in Fig.~\ref{I1_b}) can be
obtained from the known result for the pion
electromagnetic form factor as \cite{Belitsky:2001nq}
\begin{equation}
  \label{quark-kernel}
Q^{(+)}(z,x/\xi) =\Biggl\{
\mathcal{Q}\left(z,\frac{\xi+x}{2\xi}\right)-
\mathcal{Q}\left(\bar{z},\frac{\xi-x}{2\xi}\right)
\Biggr\}
+\Bigl\{z\longrightarrow \bar{z}\Bigr\},
\end{equation}
where here and in the following we use the notation $\bar{z} = 1-z$.
We have
\begin{align}
\mathcal{Q}(v,u) &= \frac{\alpha_s(\mu_R)\, C_F}{4vu}
\left(1+\frac{\alpha_s(\mu_R)}{4\pi}\, \mathcal{Q}^{(1)}(v,u)\right),
\nonumber\\[0.1em]
\mathcal{Q}^{(1)}(v,u) &=
c_1 \left[\, 2[3+\ln(vu)]\ln\left(\frac{Q^2}{\mu_F^2}\right) +
\ln^2(vu)+6\ln(vu)-\frac{\ln(v)}{\bar{v}}-\frac{\ln(u)}{\bar{u}}-
\frac{28}{3} \,\right]
\nonumber\\
& + c_2 \left[\, 2\frac{\bar vv^2+\bar u
u^2}{(v-u)^3} \bigl[\Li_2(\bar u)-\Li_2(\bar v)-\Li_2(u)+\Li_2(v)
+\ln(\bar v)\ln(u)-\ln(\bar u)\ln(v)\bigr] 
\right.
\nonumber\\[0.5em]
& \left. \qquad {}+2 \bigl[\Li_2(\bar u)+\Li_2(\bar v)
-\Li_2(u)-\Li_2(v)+\ln(\bar v)\ln(u) +\ln(\bar u)\ln(v) \bigr] \right.
\nonumber\\[0.5em]
& \left. \qquad 
+2\frac{(v+u-2vu)\ln(\bar v \bar u)}{(v-u)^2}+
4\frac{vu\ln(vu)}{(v-u)^2}-4\ln(\bar v)\ln(\bar u)
-\frac{20}{3} \,\right] 
\nonumber\\
&+ \beta_0\left[\, 
\frac{5}{3}-\ln(vu)-\ln\left(\frac{Q^2}{\mu_R^2}\right)
\,\right] \,,
\end{align}
where $\mu_R$ and $\mu_F$ denote the renormalization and factorization
scales and
\begin{align}
\beta_0 &= \frac{11}{3}N_c-\frac{2}{3}n_f , 
&
& \Li_2(z) = -\int\limits_0^z \frac{\diff t}{t}\ln(1-t) ,
\nonumber \\
c_1 &= C_F = \frac{N_c^2-1}{2N_c} ,
&
& c_2 = C_F-\frac{C_A}{2}=-\frac{1}{2N_c} .
\end{align}
Starting at NLO there is a contribution from diagrams with the
topology shown in Fig.~\ref{I1_c},
\begin{equation}
R^{(+)}(z,x/\xi) = \frac{\alpha^2_s(\mu_R)\, C_F}{8\pi z\bar z} \,
\mathcal{R}\left(z,\frac{x-\xi}{2\xi}\right) ,
\end{equation}
where
\begin{align}
\mathcal{R}(z,y) &= \Biggl\{
\frac{2y+1}{y(y+1)}\left[\frac{y}{2}\ln^2(-y)-
\frac{y+1}{2}\ln^2(y+1) \right.
\nonumber \\
& \left. \hspace{6em} {}+\bigl[ y\ln(-y)-(y+1)\ln(y+1) \bigr]
\left(\ln\left(\frac{Q^2z}{\mu_F^2}\right)-1\right)\,
\right]
\nonumber \\
& \quad {}+\frac{y\ln(-y)+(y+1)\ln(y+1)}{y(y+1)}
-\frac{V(z,y)}{y+z}
+\frac{y(y+1)+(y+z)^2}{(y+z)^2}\,W(z,y)
\Biggr\}
+\Bigl\{z\rightarrow\bar z\Bigr\}
\end{align}
with the abbreviations
\begin{align}
V(z,y) &= z\ln(-y)+\bar z\ln(y+1)+z\ln(z)+\bar z\ln(\bar z),
\nonumber \\[0.3em]
W(z,y) &= \Li_2(y+1)-\Li_2(-y)+\Li_2(z)-\Li_2(\bar z)+\ln(-y)\ln(\bar
z)-\ln(y+1)\ln(z).
\end{align}
For two-gluon exchange in the $t$-channel (Fig.~\ref{I1_a}) we can use
directly the NLO results obtained in Ref.\cite{Ivanov:2004zv},
\begin{equation}
G^{(+)}(z,x/\xi) = \mathcal{G}\left(z,\frac{x-\xi}{2\xi}\right)
\end{equation}
with
\begin{align}
\mathcal{G}(z,y) &=
\frac{\alpha_s(\mu_R)}{2z\bar{z}\, y(y+1)}\,
\left( 1+\frac{\alpha_s(\mu_R)}{4\pi}\, \mathcal{G}^{(1)}(z,y)
\right) \,,
\nonumber \\[0.3em]
\mathcal{G}^{(1)}(z,y) &= \Biggl\{
\left(\ln\left(\frac{Q^2}{\mu_F^2}\right)-1\right) \biggl[
\frac{\beta_0}{2}-\frac{2(c_1-c_2) \bigl[ y^2+(y+1)^2 \slim\bigr]
\bigl[ (y+1)\ln(y+1)-y\ln(-y) \bigr]}{y(y+1)}
\nonumber\\
& \hspace{9em} {}+\frac{c_1}{2}
\left(
\frac{y\ln(-y)}{y+1}+\frac{(y+1)\ln(y+1)}{y}
\right)+
c_1\left(\frac{3}{2}+2z\ln(\bar z)\right)
\biggr]
\nonumber\\[0.1em]
& \quad {}
-\frac{\beta_0}{2}\left(\ln\left(\frac{Q^2}{\mu_R^2}\right)-1\right)
-\frac{c_1(2y+1)V(z,y)}{2(y+z)}-\frac{3c_1-4c_2}{4}
\biggl[
\frac{y\ln^2(-y)}{y+1}+\frac{(y+1)\ln^2(y+1)}{y}
\biggr]
\nonumber\\
& \quad {}
+\bigl[\slim \ln(-y)+\ln(y+1)\bigr]
\left[c_1\left(\bar z\ln(z)-\frac{1}{4}\right)+2c_2\right]
+c_1 \bigl[\slim z\ln^2(\bar z) +(1+3z)\ln(\bar z) -2 \bigr]
\nonumber\\[0.1em]
& \quad {}-(c_1-c_2)\bigl[\slim \ln(z\bar z)-2\bigr]
\left[\frac{y\ln(-y)}{y+1}+\frac{(y+1)\ln (y+1)}{y}
\right]
\nonumber \\[0.1em]
& \quad {}+(c_1-c_2)(2y+1)\ln\left(\frac{-y}{y+1}\right)
\left[ \frac{3}{2}+\ln(z\bar z) +\ln(-y)+\ln(y+1) \right]
\nonumber \\
& \quad {}+
\Bigl(c_1 \bigl[ y(y+1)+(y+z)^2 \bigr] -c_2\slim (2y+1)(y+z)\Bigr)
\left[\,
\frac{\ln(-y)-\ln(y+1)+\ln(z)-\ln(\bar z)}{2(y+z)}
\right.
\nonumber\\
& \hspace{4em} \left.
{}-\frac{V(z,y)}{(y+z)^2}
  +\frac{y(y+1)+(y+z)^2}{(y+z)^3}\, W(z,y)
\right]
\Biggr\}+ \Bigl\{z\rightarrow\bar z\Bigr\} \,.
\end{align}
Note that in our NLO calculation we do not consider three-gluon
exchange in the $t$-channel, which in collinear factorization only
appears at NNLO in $\alpha_s$ and corresponds to odderon
exchange. Such a contribution is relevant only at very high
energies, see \cite{Hagler:2002nf}.

Since the photon has negative charge conjugation parity, the amplitude
for a pion pair produced in the $C$-even channel involves the $C$-odd
GPD combinations $F^{q(-)}$ and vice versa.  The coefficient functions
appearing in $T^{(+)}$ and $T^{(-)}$ are thus related by crossing
symmetry.  They coincide after the interchange of the $t$-channel and
the $s$-channel parton pairs in the Feynman graphs and the
corresponding interchange of the relative parton momentum fractions.
One can easily convince oneself of this relationship by comparing the
typical NLO diagrams on the left and right side in the
Fig.~\ref{fig_I}.  The prescription for the interchange of the
momentum fractions in terms of variables reads
\begin{align}
   \label{fr}
z &\leftrightarrow \frac{\xi+x}{2\xi} \,,
& 
\bar z &\leftrightarrow \frac{\xi-x}{2\xi} \,,
\end{align}
so that we have
\begin{align}
  \label{re}
Q^{(-)}\left(z,x/\xi\right)
 &= Q^{(+)}\left(\frac{\xi+x}{2\xi}, 2z-1 \right)
\, ,
\nonumber \\
R^{(-)}\left(z,x/\xi\right)
 &= R^{(+)}\left(\frac{\xi+x}{2\xi}, 2z-1 \right)
  \, ,
\nonumber \\
G^{(-)}\left(z,x/\xi\right)
 &= G^{(+)}\left(\frac{\xi+x}{2\xi}, 2z-1 \right)
  \, .
\end{align}
In \eqref{quark-kernel} to \eqref{re} we have omitted the
$i\varepsilon$ prescription, which reads $(\xi + x)/(2\xi) -
i\varepsilon$ and $(\xi-x)/(2\xi) - i\varepsilon$.
Before giving more explicit expressions for the amplitude we make a
number of simplifications:
\begin{enumerate}
\item We restrict ourselves to the asymptotic forms \eqref{2piDA-I1}
  and \eqref{2piDA-I0} for the $z$-dependence of the $2\pi$DAs.
\item We take the Omn\`es functions $f_l(s_\pi)$ and the constant $C$
  in the representations \eqref{omnes-d}, \eqref{omnes-s} to be equal
  for the quark isosinglet and for the gluon coefficients
  $\tilde{B}_{1l}^{I=0}(s_\pi)$ and $\tilde{B}_{1l}^g(s_\pi)$.  For
  the Omn\`es functions this approximation should be good at least for
  low enough $s_\pi$, where the integrands in the first lines of
  \eqref{omnes-fct-d} and \eqref{omnes-fct-s} are determined by the
  $\pi\pi$ phase shifts in the dominant integration region.  According
  to \eqref{C-chiral} we have $C \approx 1$ from chiral perturbation
  theory for both quarks and gluons.
\item We neglect the $C$-odd combination $F^{s(-)}(x,\xi,t)$ of
  nucleon GPDs.  Its forward limit $s(x) - \bar{s}(x)$ is known to be
  very small, and we assume that the same holds for finite $\xi$ and
  $t$.
\item We neglect the $2\pi$DA for strangeness, $\Phi^{s(+)}$.  Note
  that compared with $\Phi^{u(+)} = \Phi^{d(+)}$ this quantity appears
  in the amplitude \eqref{I0} with a suppression factor of either
  $F^{s(-)}(x,\xi,t)$ or $\alpha_s$.  As discussed in
  Section~\ref{sec_2piDA}, the coefficients $B_{12}^s(s_\pi)$ and
  $B_{10}^s(s_\pi)$ are small at the unphysical point $s_\pi=0$, and
  we expect that at larger $s_\pi$ they are at least not significantly
  larger than the corresponding coefficients for $u$ and $d$ quarks
  (and hence cannot compensate the suppression just mentioned).
\end{enumerate}
As a compact notation we introduce
\begin{align}
\label{I-plus}
  I_{Q}^{(+)}(\xi,t)&=
  \int_{-1}^1 \diff x \int_0^1 \diff z\,
  \varphi_0(z)\, Q^{(+)}(z,x/\xi)\, \Bigl[ 
    \tfrac{2}{3} F^{u(+)}(x,\xi,t) + 
    \tfrac{1}{3} F^{d(+)}(x,\xi,t) \Bigr] \,,
\nonumber \\
  I_{G}^{(+)}(\xi,t)&=
  \int_{-1}^1 \diff x \int_0^1 \diff z\,
  \varphi_0(z)\, G^{(+)}(z,x/\xi)\, \frac{1}{2\xi}\, F^g(x,\xi,t) \,,
\nonumber \\
  I_{R}^{(+)}(\xi,t)&=
  \int_{-1}^1 \diff x \int_0^1 \diff z\,
  \varphi_0(z)\, R^{(+)}(z,x/\xi)\, \Bigl[ 
    F^{u(+)}(x,\xi,t) + F^{d(+)}(x,\xi,t)+ F^{s(+)}(x,\xi,t)
  \Bigr]
\\
\intertext{and}
\label{I-minus}
  I_{Q}^{(-)}(\xi,t) &=
  \int_{-1}^1 \diff x \int_0^1 \diff z\,
  \varphi_1(z)\, Q^{(-)}(z,x/\xi)\, \Bigl[ 
    \tfrac{2}{3} F^{u(-)}(x,\xi,t) -
    \tfrac{1}{3} F^{d(-)}(x,\xi,t) \Bigr] \,,
\nonumber \\
  I_{G}^{(-)}(\xi,t) &=
  \int_{-1}^1 \diff x \int_0^1 \diff z\,
  \varphi_G(z)\, G^{(-)}(z,x/\xi)\, \Bigl[ 
    \tfrac{2}{3} F^{u(-)}(x,\xi,t) -
    \tfrac{1}{3} F^{d(-)}(x,\xi,t) \Bigr] \,,
\nonumber \\
  I_{R}^{(-)}(\xi,t) &=
  \int_{-1}^1 \diff x \int_0^1 \diff z\,
  \varphi_1(z)\, R^{(-)}(z,x/\xi)\, \Bigl[ 
    \tfrac{4}{3} F^{u(-)}(x,\xi,t) -
    \tfrac{2}{3} F^{d(-)}(x,\xi,t) \Bigr]
\end{align}
with
\begin{align}
  \varphi_0(z) &= z(1-z) ,
&
  \varphi_1(z) &= z(1-z)(2z-1) ,
&
  \varphi_G(z) & =z^2(1-z)^2 .
\end{align}
The integrals over $z$ can be performed analytically, whereas the $x$
integral was evaluated numerically.  The amplitudes for $\gamma^* p
\to \pi^+\pi^- p$ then take the simple form
\begin{align}
   \label{intT}
T^{(-)} &= \frac{2\pi \sqrt{4\pi\alpha}}{3 \xi Q}\,
   6 \beta F_\pi(s_\pi)\, P_1(\cos\theta)\,
   I^{(+)}(\xi,t) \,,
\nonumber \\
T^{(+)} &= \frac{2\pi \sqrt{4\pi\alpha}}{3 \xi Q}\,
   \left[ \frac{3C-\beta^2}{2}\slim f_0(s_\pi)\, P_0(\cos\theta)
      -\beta^2 f_2(s_\pi)\, P_2(\cos\theta) \right] I^{(-)}(\xi,t)
\end{align}
with
\begin{align}
  \label{intI}
I^{(+)}(\xi,t) &= 
   I_{Q}^{(+)}(\xi,t) + I_{G}^{(+)}(\xi,t) + I_{R}^{(+)}(\xi,t) \,,
\nonumber \\[0.6em]
I^{(-)}(\xi,t) &=
   - 10\slim \Bigl[ A_{20}^u(0) + A_{20}^d(0) \Bigr]
     \left( I_{Q}^{(-)}(\xi,t) + I_{R}^{(-)}(\xi,t) \right)
   - 10\slim A_{20}^g(0)\, I_{G}^{(-)}(\xi,t)
\end{align}
The corresponding expressions for $\gamma^* n \to \pi^+\pi^- n$ are
readily obtained by interchanging $F^{u(\pm)}$ and $F^{d(\pm)}$ in
\eqref{I-plus} and \eqref{I-minus}, where the quark flavor label in
the GPDs always refers to a proton target.  Explicitly we have
\begin{align}
  \label{intI-Born}
I^{(-)}(\xi,t) &= \tfrac{40}{27}\, 
  \Bigl[ A_{20}^u(0) + A_{20}^d(0) + \tfrac{3}{4} A_{20}^g(0) \Bigr]
\nonumber \\[0.2em]
 & \quad {}\times  \alpha_s(\mu_R)
  \int_{-1}^1 \diff x\, \frac{\xi^2}{(\xi+x) (\xi-x)}\,
     \Bigl[ 2 F^{u(-)}(x,\xi,t) - F^{d(-)}(x,\xi,t)
  \Bigr] + O(\alpha_s^2) \,.
\end{align}
The momentum fraction integrals in the pion fulfill $A_{20}^u(0) +
A_{20}^d(0) + \frac{3}{4} A_{20}^g(0) = 1 - A_{20}^s(0) - \frac{1}{4}
A_{20}^g(0)$, so that $I^{(-)}(\xi,t)$ depends rather weakly on the
precise values of these integrals, as was already reported in
\cite{Lehmann-Dronke:2000xq}.

In terms of the amplitude $T = T^{(-)} + T^{(+)}$, the Legendre
moments \eqref{Pl} take the form
\begin{equation}
  \label{Pl-T}
\langle P_l(\cos\theta)\rangle =
\frac{\displaystyle{
\sum\nolimits_{\mathrm{pol}} 
\int \diff y\; \diff x_B\, \diff t\, \diff s_\pi\,
     \frac{(1-y)\slim \beta}{y^{3} x_B}
     \int_{-1}^{1} \diff\cos\theta\; P_l(\cos\theta)\, |T|^2}}%
{\displaystyle{
\sum\nolimits_{\mathrm{pol}} 
\int \diff y\; \diff x_B\, \diff t\, \diff s_\pi\,
     \frac{(1-y)\slim \beta}{y^{3} x_B}
     \int_{-1}^{1} \diff\cos\theta\; |T|^2}} \,,
\end{equation}
where $\sum_{\mathrm{pol}}$ denotes summation over the polarizations
of the incoming and outgoing nucleon.  With \eqref{intT} and
\eqref{intI} we have
\begin{align}
   \label{P0}
\int_{-1}^1 \diff\cos\theta\; |T|^2 &=
   \left( \frac{2\pi\sqrt{4\pi\alpha}}{3 \xi Q} \right)^2
   \Biggl\{ \biggl(
        \frac{(3C-\beta^2)^2}{2}\, \bigl\vert f_0(s_\pi)\bigr\vert^2
      + \frac{2\beta^4}{5}\, \bigl\vert f_2(s_\pi) \bigr\vert^2
   \biggr)\, \bigl\vert I^{(-)}(\xi,t)\bigr\vert^2 
\nonumber\\
  & \hspace{7.5em}{} + 24 \beta^2\,
    \bigl\vert F_\pi(s_\pi) \bigr\vert^2\,
    \bigl\vert I^{(+)}(\xi,t) \bigr\vert^2 
  \Biggr\}
\\
\intertext{for the denominator and}
  \label{P13}
\int_{-1}^1 \diff\cos\theta\; P_1(\cos\theta)\, |T|^2 &=
   \left( \frac{2\pi\sqrt{4\pi\alpha}}{3 \xi Q} \right)^2
   12\beta\, \Re
   \biggl\{ \bigl[ F_\pi(s_\pi) \bigr]^*
      \left[\frac{3C-\beta^2}{3}\slim f_0(s_\pi)
        -\frac{4}{15}\slim \beta^2 f_2(s_\pi)
      \right]
\nonumber \\
  & \hspace{10.5em} {}\times 
    \bigl[ I^{(+)}(\xi,t) \bigr]^* I^{(-)}(\xi,t)
   \biggr\} \,,
\nonumber \\
\int_{-1}^1 \diff\cos\theta\; P_3(\cos\theta)\, |T|^2 &=
   -\left( \frac{2\pi\sqrt{4\pi\alpha}}{3 \xi Q} \right)^2
   \frac{12\beta^3}{35}\,
   \Re \biggl\{ \bigl[ F_\pi(s_\pi) \bigr]^* f_2(s_\pi)\,
       \bigl[ I^{(+)}(\xi,t) \bigr]^* I^{(-)}(\xi,t) \biggr\}
\end{align}
for the numerator of \eqref{Pl-T}.

%%%%%%%%%%%%%%%%%%%%%%%%%%%%%%%%%%%%%%%%%%%%%%%%%%%%%%%

\section{Modeling the two-pion distribution
  amplitudes\label{sec_2piModel}} 

\begin{figure}
\begin{center}
  \includegraphics[width=0.98\textwidth,clip]{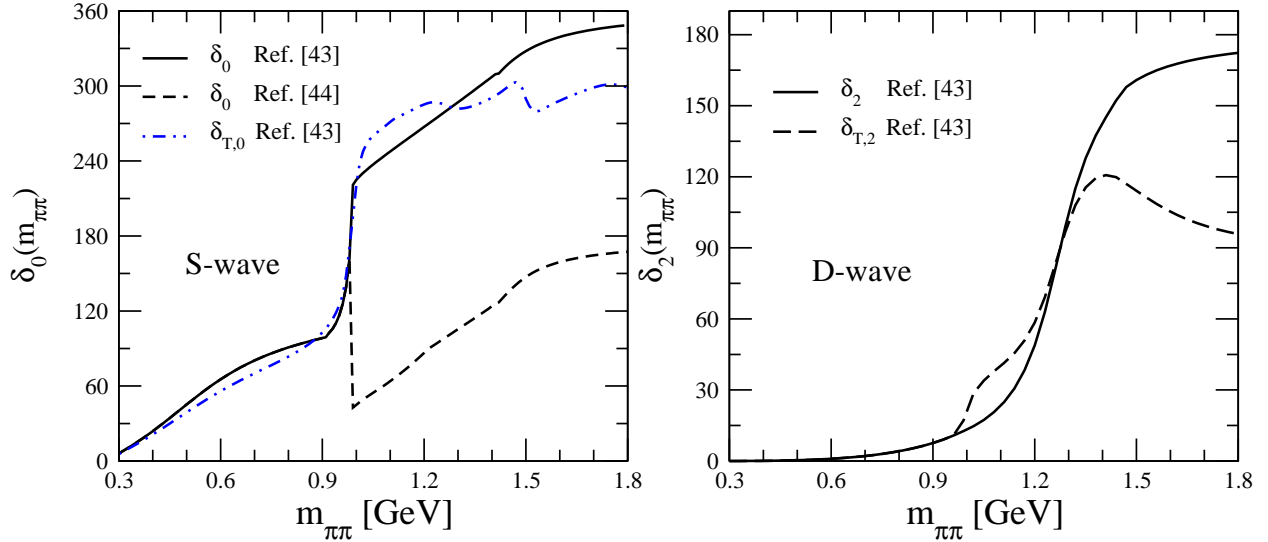}
  \caption{Phase shifts $\delta_l$ of $\pi\pi$ scattering in the
    isoscalar channel obtained by Kami{\'n}ski et al.\
    \protect\cite{Kaminski:2006yv} and by
    Bugg~\protect\cite{Bugg:2006sr}.  For the parameterization
    \protect\cite{Kaminski:2006yv} we also show the phase
    $\delta_{T,l}$ of the $\mathcal{T}$-matrix \protect\eqref{T-matrix}.
    \label{fig_Omnes_phases}}
\end{center}
\end{figure}

In Section~\ref{Omnes} we have represented the coefficients
$\tilde{B}_{10}(s_{\pi})$ and $\tilde{B}_{12}(s_{\pi})$ of the
$2\pi$DAs as 
integrals involving the phases $\tilde{\delta}_{l}(s_{\pi})$.  With
Watson's theorem \eqref{Watson} these phases are equal to the
isoscalar phase shifts $\delta_l(s_{\pi})$ from $s_{\pi}= 4m_\pi^2$ up
to the
value where $\pi\pi$ scattering in the appropriate partial wave
becomes inelastic.  Phenomenological analyses find that for
the $S$ and $D$ waves $\pi\pi$ scattering is approximately elastic up
to the two-kaon threshold at $s_{\pi}\sim 1 \gev^2$.  The $S$ wave then
becomes inelastic rather abruptly, whereas in the $D$ wave
inelasticity sets in more smoothly.  The phases of the form factors we
are interested in then differ from the corresponding phase shifts.  In
Fig.~\ref{fig_Omnes_phases} we show the $S$ and $D$ wave phase shifts
from the recent parameterization of Kami{\'n}ski et
al.~\cite{Kaminski:2006yv}, and for comparison also the result of the
analysis of the $S$ wave by Bugg~\cite{Bugg:2006sr}.

As long as a small number of channels are relevant, one can attempt an
explicit multi-channel analysis, say for the $\pi\pi$ and $K\bar{K}$
channels.  For a local operator $\mathcal{O}$ with appropriate
symmetry properties, time reversal relates the matrix element
${}_\mathrm{out}\langle \pi\pi |\slim \mathcal{O} \slim|\slim
0\rangle$ with ${}_\mathrm{in}\langle \pi\pi |\slim \mathcal{O}
\slim|\slim 0 \rangle^{*}$ and ${}_\mathrm{out}\langle K\bar{K} |\slim
\mathcal{O} \slim|\slim 0\rangle$ with ${}_\mathrm{in}\langle K\bar{K}
|\slim \mathcal{O} \slim|\slim 0 \rangle^{*}$.  In the region of $s_{\pi}$
where the scattering matrix provides a closed relation between the
states $|\pi\pi\rangle_{\mathrm{out}}$,
$|K\bar{K}\rangle_{\mathrm{out}}$ and $|\pi\pi\rangle_{\mathrm{in}}$,
$|K\bar{K}\rangle_{\mathrm{in}}$, one can then relate the phases of
the operator matrix elements with those in the $\mathcal{S}$-matrix.
Combining this information with a dispersion relation leads from the
Omn\`es representation discussed in Section~\ref{Omnes} to the
Omn\`es-Muskhelishvili problem, whose solution is considerably more
involved.  Such analyses have for instance be performed for the trace
of the energy-momentum tensor in \cite{Donoghue:1990xh} and for the
scalar quark current in
\cite{Donoghue:1990xh,Moussallam:1999aq,Liu:2000ff,Ananthanarayan:2004xy}.
We shall not attempt to do the same here for the twist-two part of the
energy-momentum tensor, which is associated with the form factors
$B_{10}(s_{\pi})$ and $B_{12}(s_{\pi})$.  Instead we wish to point out salient
results of the analysis in \cite{Ananthanarayan:2004xy}.  There it was
found that just above the $K\bar{K}$ threshold the phase of the form
factor $\Gamma(s_{\pi}) = \langle \pi\pi|\slim m_u \bar{u}u + m_d\slim
\bar{d}d \slim| 0\rangle$ starts to deviate very strongly from the $S$
wave phase shift, and instead is rather close to the phase of the
corresponding $\mathcal{T}$-matrix element in the $\pi\pi$ channel.
With the $\mathcal{S}$-matrix element in the two-pion channel
parameterized as $\eta_{\slim l} \exp[ 2i\delta_l ]$, where
$\eta_{\slim l}$ is the elasticity parameter, the corresponding
$\mathcal{T}$-matrix element is
\begin{equation}
  \label{T-matrix}
\frac{\eta_{\slim l} \exp[ 2i\delta_l ] - 1}{2i}
\end{equation}
up to a normalization factor.  Clearly, the phase $\delta_{T,l}$ of
the $\mathcal{T}$-matrix element differs from the phase shift
$\delta_l$ as soon as the elasticity deviates from $\eta_{\slim l}=1$.
This is seen in Fig.~\ref{fig_Omnes_phases}, which shows both phases
as obtained from the parameterization~\cite{Kaminski:2006yv}.

While the phase of $\Gamma(s_{\pi})$ found in \cite{Ananthanarayan:2004xy}
is well approximated by $\delta_{T,0}(s_{\pi})$, the phase of 
$\Delta(s_{\pi}) =
\langle \pi\pi|\slim m_s \bar{s}s \slim| 0\rangle$ was found to be
closer to $\delta_0(s_{\pi})$.  This difference is perhaps not too
surprising since the solution of the Omn\`es-Muskhelishvili problem
depends not only on the $\mathcal{S}$-matrix in the pion and kaon
channels but also on the relevant form factors for pions and for kaons
at the subtraction point of the dispersion relation (typically $s_{\pi}=0$).
In the present analysis we will investigate the assumptions that the
phases $\tilde{\delta}_l(s_{\pi})$ are equal to either $\delta_l(s_{\pi})$ or
$\delta_{T,l}(s_{\pi})$ for $s_{\pi}$ above the $K\bar{K}$ threshold.
We do this 
in the sense of exploring two rather extreme cases, keeping in mind
that the true phases could be far from either of them.  The phases
$\tilde{\delta}_l$ can of course be different for
$\tilde{B}_{1l}^{I=0}$ and for $\tilde{B}_{1l}^g$, but given our
simple model ansatz we take them to be the same.

The Omn\`es representations \eqref{omnes-d} and \eqref{omnes-s} depend
on the phases $\tilde{\delta}_l(s_{\pi})$ at values of $s_{\pi}$ above
the point 
where the Omn\`es functions are evaluated.  Clearly, the dependence on
large $s_{\pi}$ under the integrals is reduced for $N=2$ subtractions as
given in the second lines of \eqref{omnes-fct-d} and
\eqref{omnes-fct-s}, where the uncertainty due to the unknown behavior
of the phases at large $s_{\pi}$ is reduced at the expense of introducing an
additional subtraction constant.  In the following we shall work with
the $N=2$ Omn\`es functions, which permits a convenient estimate of
uncertainties by varying these constants.  The dynamical content of
the representations with $N=1$ and $N=2$ is of course the same, and
the simultaneous validity of the first and second lines in
\eqref{omnes-fct-d} and \eqref{omnes-fct-s} implies sum rules
\begin{alignat}{3}
  \label{omnes-sum-rules}
I_0 &= \frac{1}{\pi} \int_{4m_\pi^2}^\infty \diff s\, 
       \frac{\tilde{\delta}_0(s)}{{s}^{2}}
    & &= \frac{\diff}{\diff s_{\pi}} \ln B_{12}(0) 
       + \frac{\epsilon}{2m_\pi^2}\, \frac{B_{10}(0)}{B_{12}(0)} \,,
\nonumber \\
I_2 &= \frac{1}{\pi} \int_{4m_\pi^2}^\infty \diff s\, 
       \frac{\tilde{\delta}_2(s)}{{s}^{2}}
    & &= \frac{\diff}{\diff s_{\pi}} \ln B_{12}(0) \,.
\end{alignat}
Using the parameterization of \cite{Kaminski:2006yv} we have evaluated
the corresponding integrals in the range $4 m_\pi^2 \le s_{\pi} \le 4
m_K^2$, where the integrands are determined by the phase shifts.
Under the rather weak assumption that above the $K\bar{K}$ threshold
the phases $\tilde{\delta}_l(s_{\pi})$ remain positive (even when
dropping below $\delta_l(s_{\pi})$), this gives a lower bound on the
quantities in \eqref{omnes-sum-rules}.  We have further evaluated the
integrals with an upper cutoff at $s_{\pi} = (1.8 \gev)^2$ and up to
$s_{\pi}=\infty$, assuming either $\tilde{\delta}_l = \delta_l$ or
$\tilde{\delta}_l = \delta_{T,l}$.  The results are collected in
Table~\ref{tab:integrals} and provide an estimate of the possible
contribution to the integrals \eqref{omnes-sum-rules} from the region
$s_{\pi} \ge 4m_K^2$.  For the $S$ wave the contribution from $s_{\pi}
\le 4m_K^2$ gives an important part of the total result.  In contrast,
the $D$ wave is strongly suppressed in that region, and practically
the entire contribution to the integral $I_2$ comes from $s_{\pi}$
above the two-kaon threshold.
We note that the instanton model calculation reported in
\cite{Polyakov:1998ze} obtained
\begin{align}
  \label{poly-model-2}
 \frac{\diff}{\diff s_{\pi}} \ln{B^{I=0}_{12}}(0)
&=\frac{\diff}{\diff s_{\pi}} \ln{B^{I=0}_{10}}(0)
 =\frac{N_c}{3}\, \frac{1}{(4\pi f_\pi)^2} \approx 0.73 \gev^{-2} \,,
\\
\intertext{which together with the relation \protect\eqref{B12-10}
  from chiral perturbation theory gives}
\label{poly-model-0}
\frac{\diff}{\diff s_{\pi}} \ln B_{12}^{I=0}(0) 
  + \frac{\epsilon}{2m_\pi^2}\, \frac{B_{10}^{I=0}(0)}{B_{12}^{I=0}(0)}
&\approx \frac{\diff}{\diff s_{\pi}} \ln B_{10}^{I=0}(0) 
         - \frac{19}{60}\, \frac{1}{(4\pi f_\pi)^2}
\approx 0.50 \gev^{-2} \,.
\end{align}
For $\frac{\diff}{\diff s_{\pi}} \ln B_{12}^{I=0}(0)$ the value in
\eqref{poly-model-2} is not very different from our estimates for
$I_2$ in Table~\ref{tab:integrals}, but for $I_0$ even our result from
the region $s_{\pi} \le 4m_K^2$ is significantly higher than
\eqref{poly-model-0}.  Since in that region the pion phase shifts are
reasonably well known and equal to $\tilde{\delta}_0$, we must
conclude that the result for $\frac{\diff}{\diff s_{\pi}} \ln
B_{10}^{I=0}(0)$ in \cite{Polyakov:1998ze} is implausibly small---it
would only be consistent with the sum rule \eqref{omnes-sum-rules} if
$\tilde{\delta}_0$ became significantly negative for $s_{\pi} \ge 4m_K^2$ or
if the relation \eqref{B12-10} from one-loop chiral perturbation
theory were invalidated by huge corrections from higher orders.

\begin{table}
  \caption{\label{tab:integrals} The integrals $I_0$ (left) and $I_2$
    (right) defined in \protect\eqref{omnes-sum-rules}, evaluated with
    different upper cutoffs $s_{\mathrm{max}}$ on $s_{\pi}$.  The phases
    $\delta_l(s_{\pi})$ and $\delta_{T,l}(s_{\pi})$ are taken from the
    parameterization in \protect\cite{Kaminski:2006yv}.  All integrals
    are given in units of $\gev^{-2}$.}
\begin{center}
\renewcommand{\arraystretch}{1.2}
\begin{tabular}{lccc} \hline
  & \multicolumn{3}{c}{$s_{\mathrm{max}}$} \\
  & $4 m_K^2$ & $(1.8 \gev)^2$ & $\infty$ \\ \hline
$\tilde{\delta}_0 = \delta_0 \rule{0pt}{1.2em}$ &  
  $2.02$ & $3.13$ & $3.74$ \\
$\tilde{\delta}_0 = \delta_{T,0}$ & 
  $2.02$ & $2.34$ & $2.55$ \\ \hline
\end{tabular}
\hspace{2em}
\begin{tabular}{lccc} \hline
  & \multicolumn{3}{c}{$s_{\mathrm{max}}$} \\
  & $4 m_K^2$ & $(1.8 \gev)^2$ & $\infty$ \\ \hline
$\tilde{\delta}_2 = \delta_2 \rule{0pt}{1.2em}$ &  
  $0.04$ & $0.37$ & $0.67$ \\
$\tilde{\delta}_2 = \delta_{T,2}$ & 
  $0.04$ & $0.33$ & $0.48$ \\ \hline
\end{tabular}
\end{center}
\end{table}

\begin{figure}[t]
\begin{center}
\includegraphics[width=0.98\textwidth,clip]{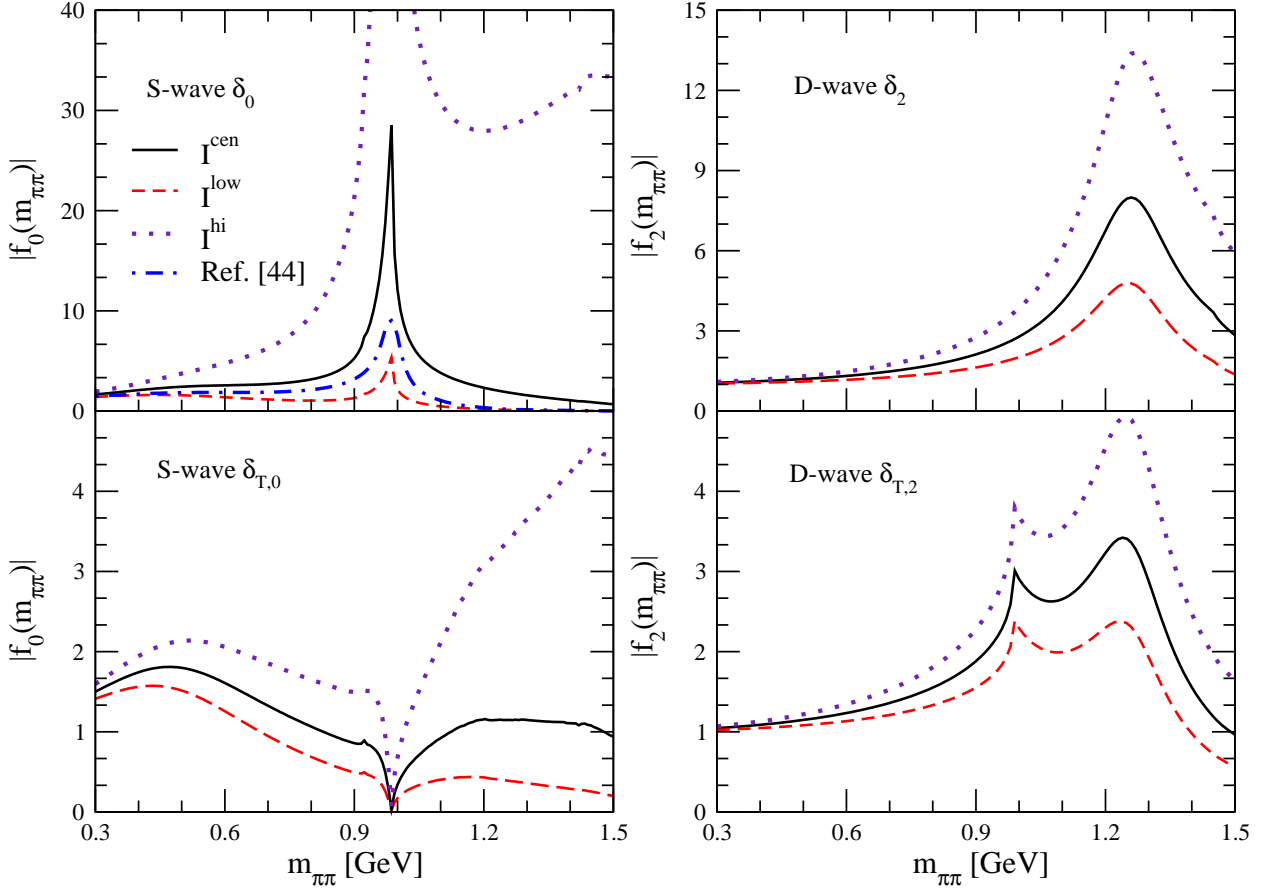}
\caption{\label{fig_Omnes_var} Absolute values of the Omn\`es
  functions for $N=2$ subtractions, assuming $\tilde{\delta}_l =
  \delta_l$ (top) or $\tilde{\delta}_l = \delta_{T,l}$ (bottom).  The
  different values taken for the subtraction constants are explained
  in the text.}
\end{center}
\end{figure}

In Fig.~\ref{fig_Omnes_var} we show the absolute values of the Omn\`es
functions evaluated with $N=2$ subtractions, obtained with either
$\tilde{\delta}_l = \delta_l$ or $\tilde{\delta}_l = \delta_{T,l}$
taken from the parameterization in \cite{Kaminski:2006yv}.  To explore
the dependence on the subtraction constants, we take as central values
$I_l^{\mathrm{cen}}$ those obtained with $s_{\mathrm{max}}=\infty$ in
Table~\ref{tab:integrals}.  For the $S$ wave we take as a lowest value
$I_0^{\mathrm{low}}$ the one obtained with $s_{\mathrm{max}}= 4m_K^2$
and as highest value $I_0^{\mathrm{hi}} = I_0^{\mathrm{cen}} +
(I_0^{\mathrm{cen}} - I_0^{\mathrm{low}})$.  For the $D$ wave we take
instead $I_2^{\mathrm{low}} = 0.5\, I_2^{\mathrm{cen}}$ and
$I_2^{\mathrm{hi}} = 1.5\, I_2^{\mathrm{cen}}$.
For the $S$ wave and the assumption $\tilde{\delta}_0 = \delta_0$ we
also show the result from the parameterization in \cite{Bugg:2006sr}.
Here the integrals in both the Omn\`es function \eqref{omnes-fct-s}
and in the corresponding subtraction constant \eqref{omnes-sum-rules}
are taken with an upper cutoff $s_{\text{max}} = (1.8 \gev)^2$ since
we do not have an analytic parameterization up to $s_{\pi}=\infty$ in
this case.  A corresponding truncation of the integrals with the
parameterization of \cite{Kaminski:2006yv} has only a moderate effect,
and we do not show the corresponding curve.

We see that the values of the subtraction constants have a visible
effect on the Omn\`es functions, as well as the choice of
parameterization for the $\pi\pi$ phase shifts and elasticity
parameters.  The largest uncertainty in the Omn\`es functions is
however due to the values of $\tilde{\delta}_l(s_{\pi})$ for $s_{\pi}$
above the 
kaon threshold, as exemplified by the two assumptions
$\tilde{\delta}_l = \delta_l$ and $\tilde{\delta}_l = \delta_{T,l}$.
Whereas the former produces a clear peak of $|f_0(s_{\pi})|$ around $s_{\pi}=
1\gev^2$, whose height depends on further details, the latter gives a
dip at the same position.  For $|f_2(s_{\pi})|$ the differences are less
extreme but still significant.  In the next section we will compare
the consequences of these Omn\`es functions on the two-pion mass
spectrum and angular distribution observed at HERMES.

%%%%%%%%%%%%%%%%%%%%%%%%%%%%%%%%%%%%%%%%%%%%%%%%%%%%%%%

\section{Results for two-pion electroproduction\label{sec_results}}

We have now all ingredients necessary to evaluate the invariant mass
spectrum and angular distribution of the two pions produced in
$\gamma^* + N\to \pi^+\pi^- + N$.  We take the factorization and
renormalization scales as $\mu_F = \mu_R = Q$ in the hard-scattering
formulae.  For the nucleon GPDs we use the model described in
Section~\ref{sec_gpdmodel}.  The $2\pi$DAs are calculated within the
model specified above \eqref{I-plus}, based on the asymptotic
$z$-dependence and the Omn\`es representations developed in Section 
\ref{Omnes}.  This leads to the expressions \eqref{intT} and
\eqref{intI} for the scattering amplitude.  We take $C=1$ for the
constant in \eqref{omnes-s} and expect that the chiral corrections of
order $m_\pi^2 /\Lambda_\chi^2$ to this quantity have a negligible
effect on our results, given the other uncertainties we have to deal
with.  The Omn\`es functions are calculated with $N=2$ subtractions,
using the phases and subtraction constants presented in the previous
section.  For the quark and gluon momentum fractions appearing in
\eqref{intI} we take $A_{20}^u(0) + A_{20}^d(0) = A_{20}^g(0) = 0.5$,
in line with the parton densities of the pion at moderate
factorization scales \cite{Gluck:1999xe}.  A change of these values
has only little effect on our results, as explained after
\eqref{intI-Born}.

For the timelike pion form factor $F^{}_\pi(m_{\pi\pi}^2)$ we take the
parameterization given in \cite{Melikhov:2003hs}, which is in good
agreement with data from $e^+e^-\to \pi^+\pi^-$ and also with the pion
phase shift $\delta_1$ in the $P$ wave.  We note that according to the
analysis in \cite{Eidelman:2003uh}, inelasticity has only a small
effect on the difference between $\delta^{}_1(m_{\pi\pi}^2)$ and the
phase of $F^{}_\pi(m_{\pi\pi}^2)$ for $m_{\pi\pi} \lsim 1.3 \gev$.
Our results are almost unchanged if we use the parameterizations for
$F^{}_\pi(m_{\pi\pi}^2)$ from \cite{Pich:2001pj} or
\cite{Guerrero:1997ku} instead of \cite{Melikhov:2003hs}.

When giving results for a deuteron target we will assume that the
production process is incoherent, $\gamma^* + d \to \pi^+\pi^- + p +
n$, with scattering either on the proton or the neutron. For the
average $t = -0.29 \gev^2$ of the HERMES measurement
\cite{Airapetian:2004sy} this should be a good approximation since
elastic scattering on the deuteron is strongly suppressed at that
value of $t$.  In addition we neglect nuclear effects and treat proton
and neutron as quasi-free.  We then simply have
\begin{align}
\diff\sigma(\gamma^* d) 
  &= \diff\sigma(\gamma^* p) + \diff\sigma(\gamma^* n) \,,
&
\langle P_l \rangle_d^{} &=
 \frac{\diff\sigma_p}{\diff\sigma_p 
     + \diff\sigma_n}\, \langle P_l \rangle_p^{}
+\frac{\diff\sigma_n}{\diff\sigma_p 
     + \diff\sigma_n}\, \langle P_l \rangle_n^{} \,,
\end{align}
where subscripts $d$, $p$, $n$ refer to the different targets.  In the
following we give results for two kinematic points, which correspond
to the average kinematics of the HERMES measurement
\cite{Airapetian:2004sy}:
\begin{align}
  \label{kin-p}
t &= -0.43 \gev^2, & Q^2 &= 3.2 \gev^2, & x_B &= 0.16 & & \text{for a
  proton target,}
\\
  \label{kin-d}
t &= -0.29 \gev^2, & Q^2 &= 3.3 \gev^2, & x_B &= 0.16 & & \text{for a
  deuteron target.}
\end{align}

\begin{figure}
\begin{center}
\includegraphics[width=0.67\textwidth, clip]{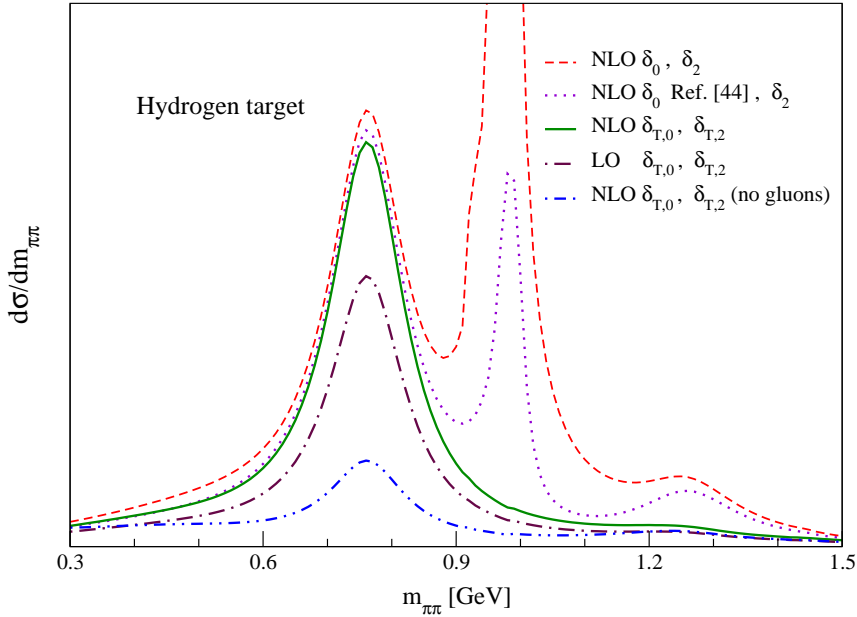}
\caption{\label{fig_thcounts} Two-pion invariant mass spectrum (in
  arbitrary units) for $\gamma^* + p\to \pi^+\pi^- + p$, calculated
  with different assumptions for the Omn\`es functions as explained in
  the text.  Phases are obtained from
  Ref.~\protect\cite{Kaminski:2006yv} unless explicitly indicated.
  The plot is for the average kinematics \protect\eqref{kin-p} of the
  HERMES measurement~\protect\cite{Airapetian:2004sy}.}
\end{center}
\end{figure}

Our results for the invariant mass spectrum of two pions produced from
a hydrogen target are given in Fig.~\ref{fig_thcounts}.  We show them
for the Omn\`es functions calculated from the parameterizations of
Kami{\'n}ski et al.\ \cite{Kaminski:2006yv} and of
Bugg~\cite{Bugg:2006sr} with the hypothesis $\tilde{\delta}_l =
\delta_l$, and in the case of \cite{Kaminski:2006yv} also for
$\tilde{\delta}_l = \delta_{T,l}$.  In all cases, the central values
$I_l^\mathrm{cen}$ for the subtraction constants have been used.  As
expected from our discussion of the Omn\`es functions, the ansatz
$\tilde{\delta}_0 = \delta_0$ produces a clearly visible peak in the
mass spectrum around $m_{\pi\pi} = 1 \gev$, although its height
strongly depends on the phase shifts used.  The HERMES measurement
\cite{Airapetian:2004sy} did not find any indication of such a
pronounced peak, and we conclude that these data strongly disfavor the
hypothesis that the phase of $\tilde{B}_{10}$ is given by the $S$ wave
phase shift above the two-kaon threshold.  In the following we will
therefore restrict the discussion to our alternative hypothesis
$\tilde{\delta}_0 = \delta_{T,0}$.  In the case of the $D$ wave the
assumption $\tilde{\delta}_2 = \delta_2$ produces a peak around the
mass of the $f_2(1270)$.  It is however much less pronounced than the
one in the $S$ wave, and we find that the invariant mass spectrum
shown in \cite{Airapetian:2004sy} does not allow a strong conclusion
on the phase of $\tilde{B}_{12}$.  The same discussion applies for a
deuterium target, i.e., the assumption $\tilde{\delta}_0 = \delta_{0}$
produces a clear mass peak around $m_{\pi\pi} = 1 \gev$, which is not
seen in the data, whereas the mild peak around $m_{\pi\pi} = 1.27 \gev$
produced by $\tilde{\delta}_2 = \delta_2$ cannot be ruled out by the
data.

We also show in Fig.~\ref{fig_thcounts} the result of taking the LO
approximation for the hard-scattering subprocess (for one choice of
Omn\`es functions).  The effect of the NLO corrections on this
observable is clearly visible, but not of a size which would make us
worry about the stability of the perturbative expansion.  Note that we
are not giving the absolute size of the cross section here: on one
hand there is no experimental measurement to compare with, and on the
other hand we expect important power corrections to our leading-twist
calculation of this observable, as discussed in
Section~\ref{sec:fact}.
The lowest curve in Fig.~\ref{fig_thcounts} shows the result we obtain
when setting the gluon GPD to zero.  We see that even in HERMES
kinematics there is a substantial contribution from gluon exchange to
the $P$ wave production amplitude, confirming similar findings in
\cite{Diehl:2004wj,Goloskokov:2006hr}.

\begin{figure}
\begin{center}
\includegraphics[width=0.78\textwidth,%
clip=true]{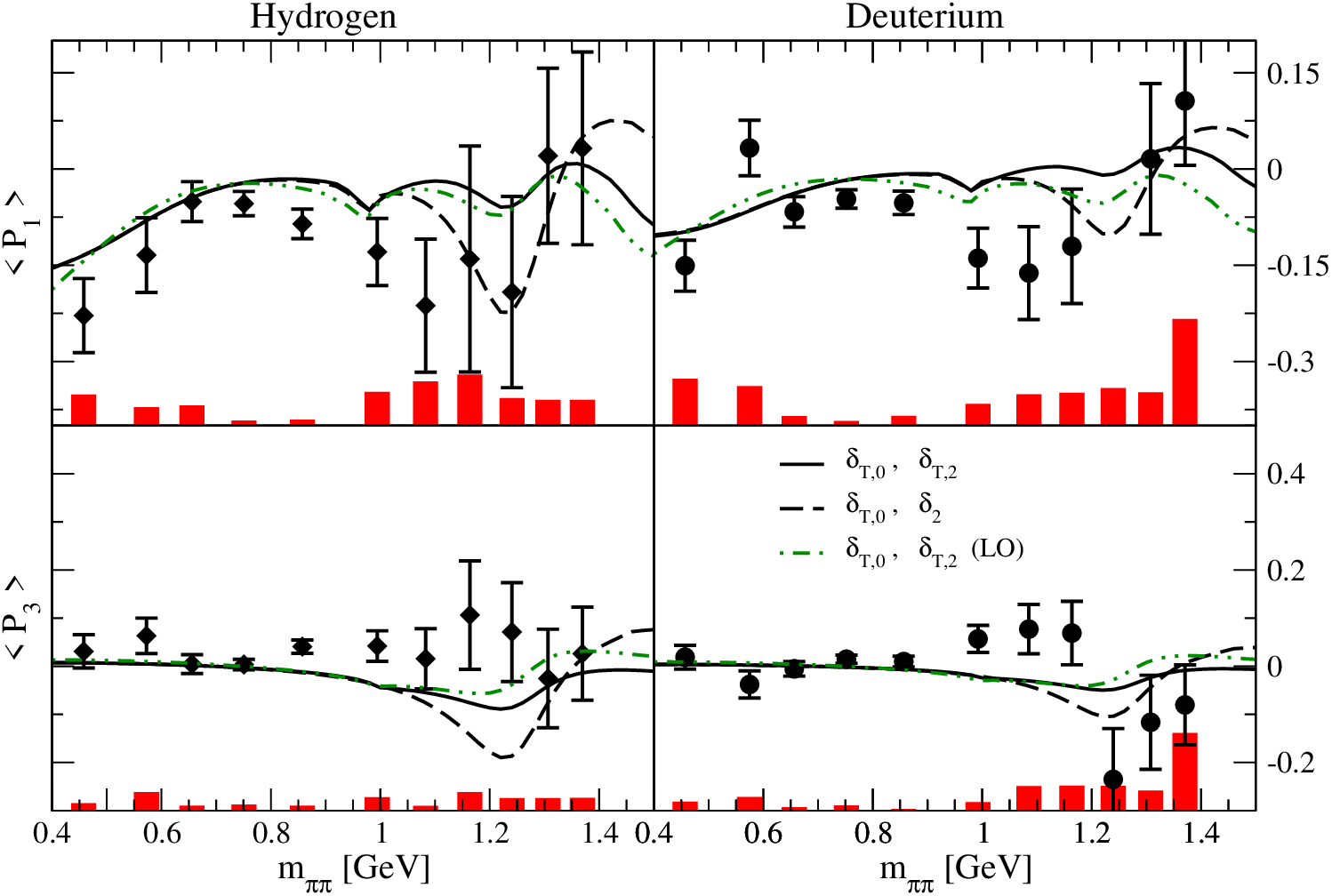}
\caption{\label{fig_mom} Legendre moments $\langle P_1\rangle$ and
  $\langle P_3\rangle$ for a hydrogen and deuterium target.  The
  curves are calculated with the models specified in the text for the
  average kinematics \protect\eqref{kin-p} and \protect\eqref{kin-d}
  of the HERMES data \protect\cite{Airapetian:2004sy}.  The systematic
  uncertainty of the measurement is represented by the histograms.}
\end{center}
%\end{figure}
%
%\begin{figure}[p]
\begin{center}
\includegraphics[width=0.78\textwidth,%
clip=true]{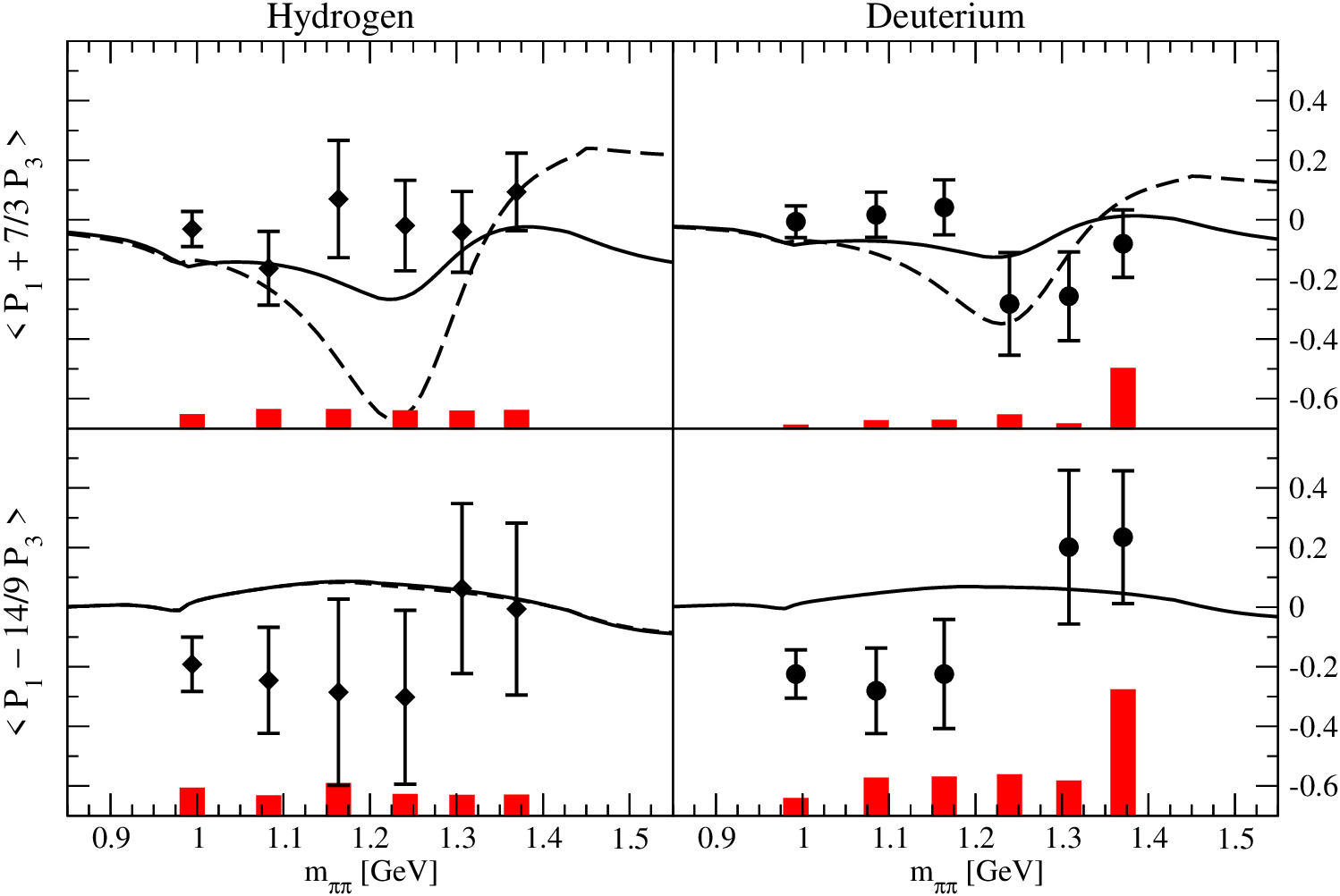}
\caption{\label{fig_momcomb} Data and theory for the linear
  combinations \protect\eqref{l_comb_0} and \protect\eqref{l_comb_1}
  of Legendre moments, obtained with the same Omn\`es functions as in
  Fig.~\protect\ref{fig_mom}.  Curves here and in the following
  figures are for NLO.}
\end{center}
\end{figure}

In Fig.~\ref{fig_mom} we show our results for the Legendre moments
\eqref{Pl} as a function of $m_{\pi\pi}$.  Here and in the following
figures we always use the phases
from \cite{Kaminski:2006yv}.  Given the experimental errors we find
the overall agreement between data and theory fair, although clearly
not perfect.  The two curves correspond to the hypotheses
$\tilde{\delta}_2 = \delta_{2}$ or $\tilde{\delta}_2 = \delta_{T,2}$.
Whereas at face value the former hypothesis seems to be preferred by
the data on $\langle P_1(\cos\theta) \rangle$, the opposite holds for
$\langle P_3(\cos\theta) \rangle$.
In Fig.~\ref{fig_momcomb} we compare the same two scenarios for the
linear combinations \eqref{l_comb_0} and \eqref{l_comb_1} of Legendre
moments at higher $m_{\pi\pi}$ values.  The combination $\langle P_1 +
\frac{7}{3} P_3\rangle$ is only sensitive to amplitudes with total
helicity $\lambda=0$ of the two pions, which are those we can
calculate using the factorization theorem.  For this observable, the
curve obtained with $\tilde{\delta}_2 = \delta_{2}$ is rather
disfavored by the hydrogen data.  The combination $\langle P_1 -
\frac{14}{9} P_3\rangle$ is sensitive to a $\lambda=0$ contribution
from the $S$ wave, which comes out rather small in this $m_{\pi\pi}$
range, and to $\lambda = \pm 1$ in the $D$ wave, which is absent in
our leading-twist calculation (so that there is no difference between
the two model curves).
We also show in Fig.~\ref{fig_mom} the result obtained at LO in
$\alpha_s$ for one choice of the Omn\`es functions.  The effect of the
NLO corrections is again found to be moderate but not negligible.

\begin{figure}[p]
\begin{center}
\includegraphics[width=0.82\textwidth,%
clip=true]{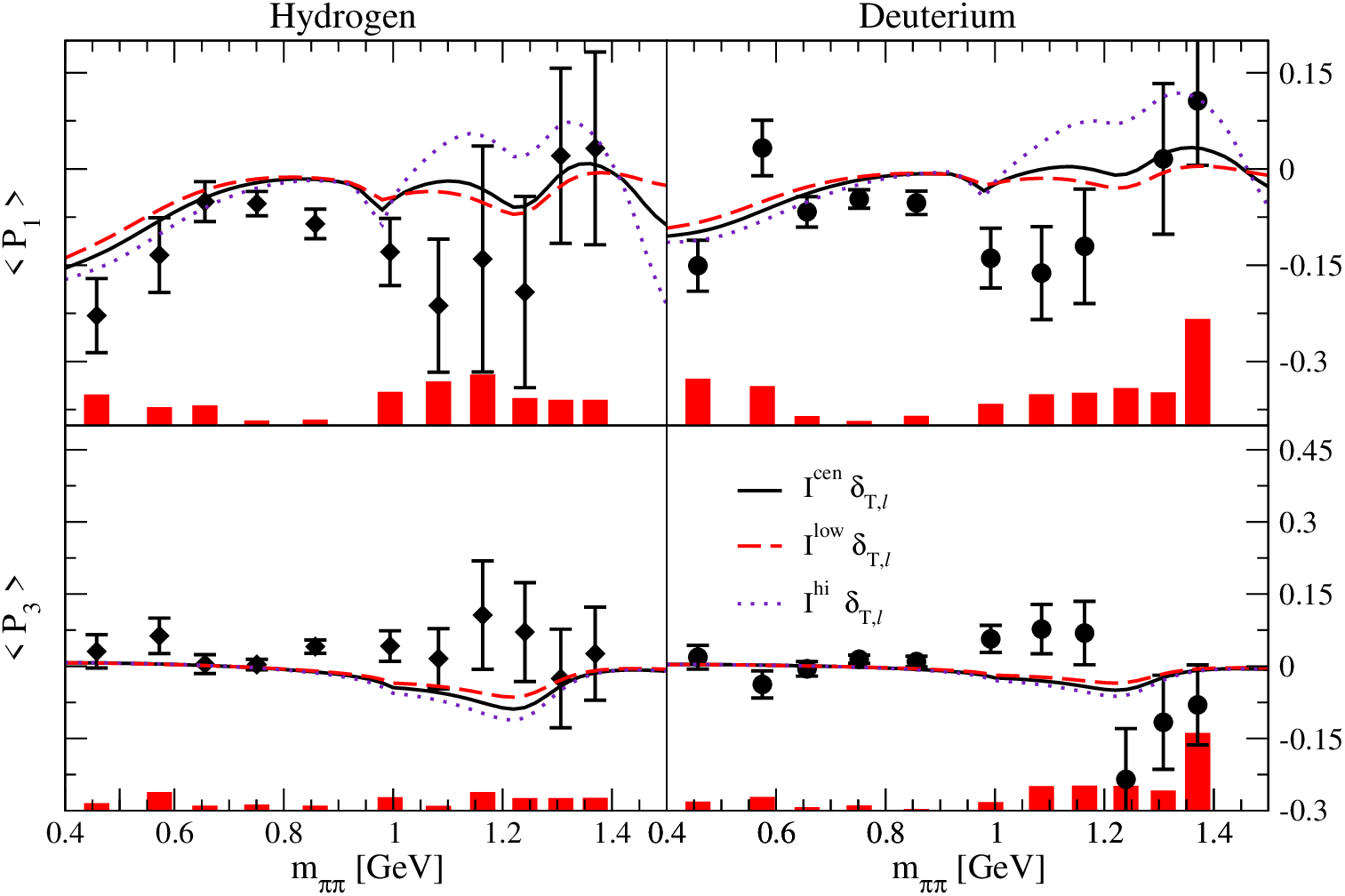}
\caption{\label{fig_momvar} As Fig.~\protect\ref{fig_mom} but for
  Omn\`es functions calculated with $\tilde{\delta}_l = \delta_{T,l}$
  and our high, central,
  or low estimates for the subtraction constants $I_l$ specified in
  Sect.~\protect\ref{sec_2piModel}.}
\end{center}
\begin{center}
\includegraphics[width=0.82\textwidth,%
clip=true]{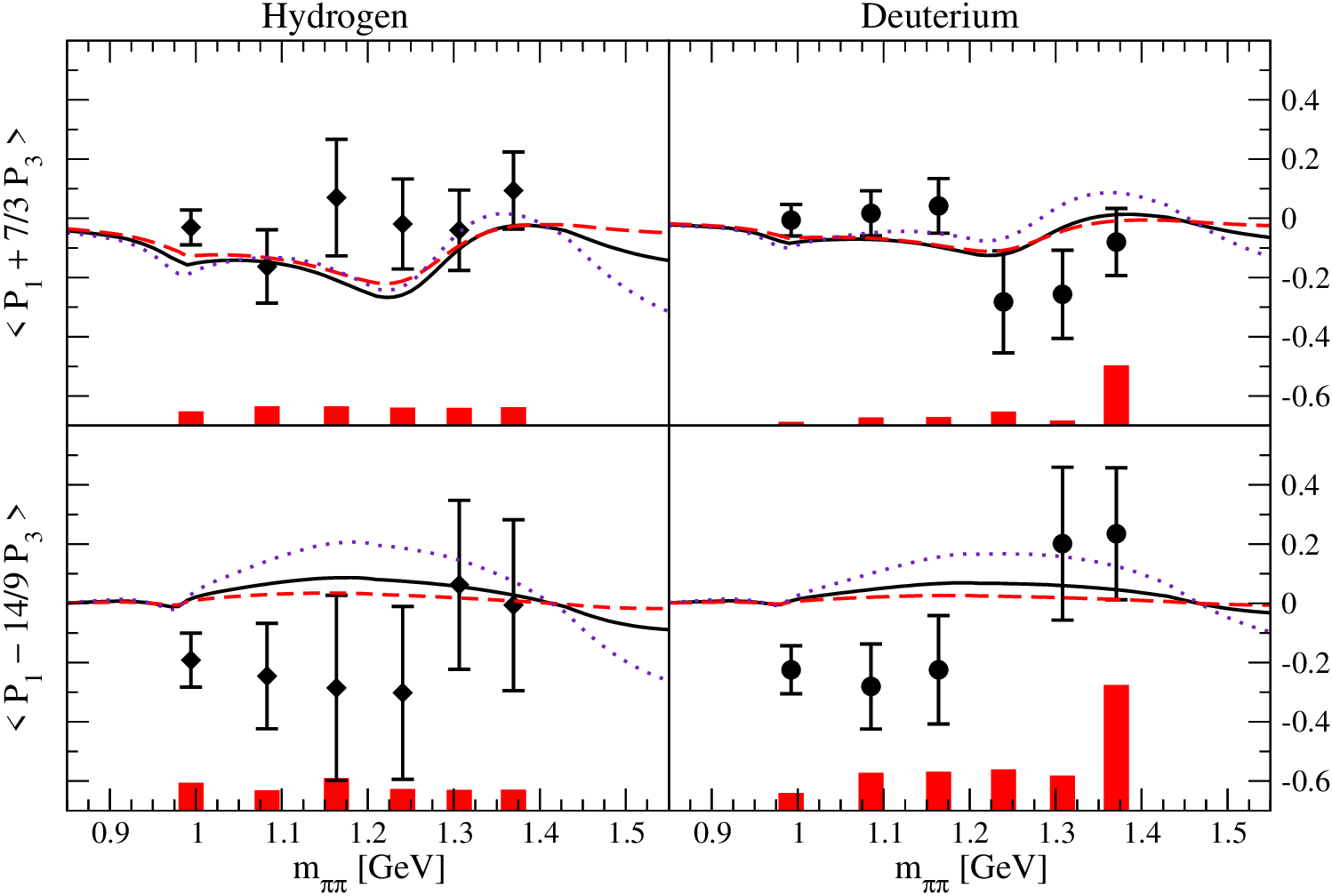}
\caption{\label{fig_momcombvar} As Fig.~\protect\ref{fig_momvar} but
  for the linear combinations \protect\eqref{l_comb_0} and
  \protect\eqref{l_comb_1} of Legendre moments.}
\end{center}
\end{figure}

Figures~\ref{fig_momvar} and \ref{fig_momcombvar} show our results
obtained with $\tilde{\delta}_l = \delta_{T,l}$ in both $S$ and $D$
waves as a function of the subtraction constants $I_0$ and $I_2$
discussed in Section~\ref{sec_2piModel}, where we have taken the high,
central, or low values for both constants at a time.  We see that the
impact of these values on the observables is not negligible but beyond
the accuracy of the presently available data.

\begin{figure}
\begin{center}
\includegraphics[width=0.98\textwidth,clip]{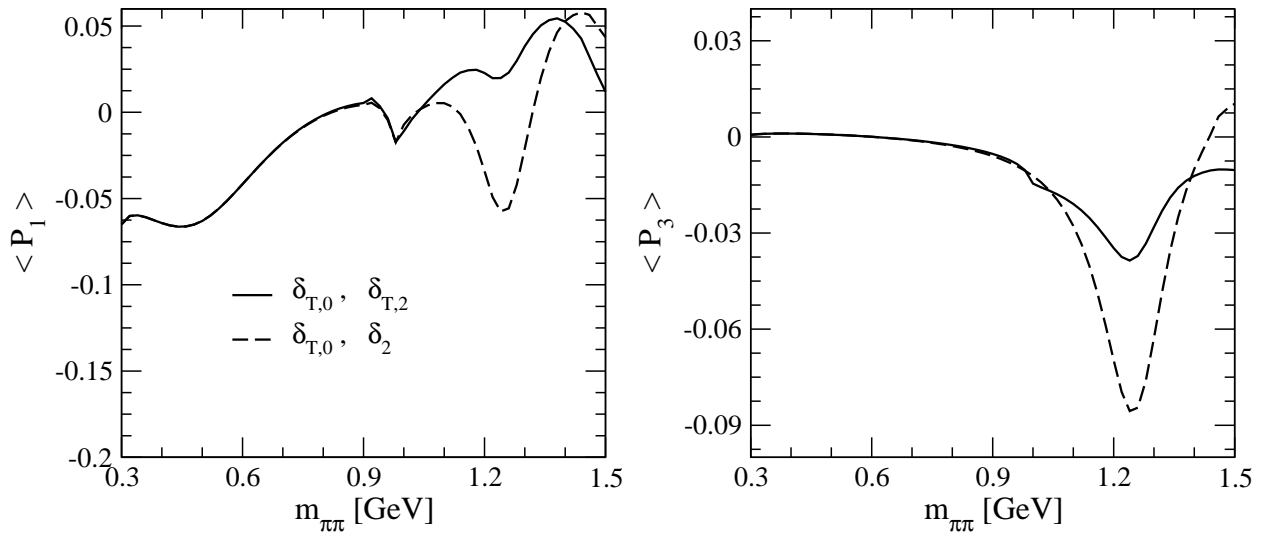}
\caption{\label{fig_Q7_exp_mpp} Predictions for the Legendre moments
  for a deuterium target in typical kinematics of the COMPASS
  experiment, $x_B = 0.08$, $t= -0.27 \gev^2$ and $Q^2 = 7 \gev^2$.
  The Omn\`es functions used are the same as in
  Fig.~\protect\ref{fig_mom}.}
\end{center}
\end{figure}

We finally show in Fig.~\ref{fig_Q7_exp_mpp} predictions for the
Legendre moments in kinematics typical of the COMPASS experiment.  In
this case one can afford higher values of $Q^2$, which increases the
reliability of the leading-twist approach we use here.  We see that
compared with the HERMES kinematics the overall size of the Legendre
moments is decreased.  This is not too surprising, since the
production of two pions in the $P$ wave is enhanced by the growth of
the gluon distribution with decreasing $x_B$, and the Legendre moments
reflect the interference of the $P$ wave with the $S$ or $D$ waves
(which are insensitive to the gluon distribution).  Comparing with the
LO results, we find that the quantitative effect of NLO corrections on
the invariant mass spectrum and on the Legendre moments in the
kinematics of Fig.~\ref{fig_mom} is of similar size as in the HERMES
case shown above.

%%%%%%%%%%%%%%%%%%%%%%%%%%%%%%%%%%%%%%%%%%%%%%%%%%%%%%%

\section{Summary\label{sec_sum}}

We have calculated exclusive electroproduction of $\pi^+\pi^-$ pairs
on the nucleon at NLO in $\alpha_s$, focusing on the kinematics of the
existing measurement at HERMES and of a possible analysis at COMPASS.
We find that the effects of NLO corrections on the invariant mass
spectrum are moderate, although not negligible.  The same holds for
NLO effects on the angular distribution of the produced pions,
quantified by the Legendre moments \eqref{Pl}.  This indicates that
the perturbative expansion is well behaved for the process in the
kinematics we have studied.

In addition to generalized parton distributions, which appear in a
number of hard exclusive processes, a crucial ingredient for the
description of our reaction are two-pion distribution amplitudes,
which describe the exclusive hadronization of a parton pair into
$\pi^+\pi^-$.  The lowest moments of these distribution amplitudes are
form factors of the energy-momentum tensor.  We have examined in
detail their representation as dispersion integrals, which requires
special care because of the presence of both $S$ and $D$ wave
components.  The behavior of these form factors directly influences
the invariant mass and angular distribution of the pion pair in
electroproduction.  A salient feature of the HERMES measurement
\cite{Airapetian:2004sy} is the absence of a clear peak in the $S$
wave for $m_{\pi\pi}$ around the mass of the $f_0(980)$.  We
take this as a strong indication that, once the two-kaon channel is
opened, the phase of the relevant form factors differs strongly from
the two-pion phase shift $\delta_0$.  More detailed investigation of
the angular distribution suggests that a similar statement may hold in
the $D$ wave for $m_{\pi\pi}$ above the mass of the $f_2(1270)$.

Considering two simple hypotheses for the phase of the two-pion
distribution amplitudes we have seen that the range of model
predictions is far greater than explored so far in the literature, and
that the existing data can distinguish between different model
assumptions.  A more sophisticated treatment would be a two-channel
analysis, similar to what has been done for the form factors of the
scalar quark current
\cite{Donoghue:1990xh,Moussallam:1999aq,Liu:2000ff,Ananthanarayan:2004xy}.
Such an analysis would involve further low-energy constants, and in
our opinion would greatly benefit from further data with smaller
statistical errors and, preferably, at higher values of $Q^2$.

In conclusion we find that pion electroproduction is a case for which
on one hand higher-order QCD corrections are under control and on the
other hand chiral perturbation theory is rather advanced and well
understood.  This process may therefore be seen as a show case for the
interplay of both approaches.  Further progress can come from the
improved knowledge of GPDs and GDAs, from extended calculations in
chiral perturbation theory, and from experimental analysis of the
two-pion system above the kaon threshold.

%%%%%%%%%%%%%%%%%%%%%%%%%%%%%%%%%%%%%%%%%%%%%%%%%%%%%%%

\section*{Acknowledgments}

We are grateful to J.~Gasser, Ph.~H\"agler, H. Leutwyler, A.~Manashov
and D. M\"uller for helpful discussions, and to R.~Fabbri and
F.-H.~Heinsius for providing details on the HERMES and COMPASS
experiments, respectively.  We thank D. Bugg for providing the values
of the pion phase shift and elasticity parameter in
\cite{Bugg:2006sr}.  This work has been supported by BMBF, contract
number 06RY258, and by the Helmholtz Association, contract number
VH-NG-004.  The work of D.I.~was also partially supported by grants
RFBR-06-02-16064 and NSh 5362.2006.2.

%%%%%%%%%%%%%%%%%%%%%%%%%%%%%%%%%%%%%%%%%%%%%%%%%%%%%%%%%%%%%

\end{document}